\documentclass[
  aps,
  prd,
  reprint,
  twocolumn,
  groupedaddress,
  floatfix,
  nofootinbib
]{revtex4-2}

\usepackage{amsmath,amssymb}
\usepackage{amsfonts}
\usepackage{mathtools}
\usepackage{bm}
\usepackage{bbm, dsfont}
\usepackage{units}
\usepackage{tabularx}
\usepackage{extarrows}
\usepackage{enumitem}
\usepackage{tikz}

\usepackage{hyperref}
\hypersetup{
    colorlinks,
    linkcolor={blue},
    citecolor={green!70!black},
    urlcolor={blue!80!black},
    pdftitle={SU(3) LECs},
    pdfauthor={RQCD},
}

\newcommand{\GeV}{\text{GeV}}
\newcommand{\MeV}{\text{MeV}}	
\newcommand{\fm}{\text{fm}}
\DeclareMathOperator{\tr}{tr}

\newcommand{\dd}{\mathrm{d}}
\DeclareRobustCommand{\bmr}[1]{\bm{#1}}
\newcommand{\mydot}[1]{\tikz{\filldraw[draw=#1,fill=#1] (0,0) circle (3pt);}\ \ }
\newcommand{\mydiamond}[1]{\tikz{\filldraw[draw=#1,fill=#1,rotate around={45:(2.2pt,2.2pt)}] (0,0) rectangle (4.4pt,4.4pt);}\ \ }
\newcommand{\myhexagon}[1]{\tikz{\filldraw[draw=#1,fill=#1,] (0:3pt) \foreach \x in {60,120,...,360} {  -- (\x:3pt)};}\ \ }

\definecolor{A652}{HTML}{4bb062}
\definecolor{A653}{HTML}{278f48}
\definecolor{A650}{HTML}{026f2e}
\definecolor{H101}{HTML}{f14432}
\definecolor{U103}{HTML}{d92523}
\definecolor{rqcd021}{HTML}{bc141a}
\definecolor{rqcd017}{HTML}{980c13}
\definecolor{B450}{HTML}{4a98c9}
\definecolor{rqcd030}{HTML}{2575b7}
\definecolor{X450}{HTML}{0a539e}
\definecolor{N202}{HTML}{8683bd}
\definecolor{X250}{HTML}{6d57a6}
\definecolor{X251}{HTML}{552a90}
\definecolor{N300}{HTML}{ffb100}
\definecolor{J500}{HTML}{7a7a7a}

\begin{document}

\title{Leading order mesonic and baryonic SU(3) low energy constants from $\bmr{N_f = 3}$~lattice~QCD}

\author{Gunnar~S.~Bali}
\email[]{gunnar.bali@ur.de}
\altaffiliation{%
Department of Theoretical Physics, 
Tata Institute of Fundamental Research, 
Homi Bhabha Road, Mumbai 400005, India.
}

\author{Sara~Collins}
\email[]{sara.collins@ur.de}

\author{Wolfgang~S\"oldner}
\email[]{wolfgang.soeldner@ur.de}

\author{Simon~Weish\"aupl}
\email[]{simon.weishaeupl@ur.de}

\affiliation{%
Institut f\"ur Theoretische Physik, 
Universit\"at Regensburg, 
93040 Regensburg, Germany.
}

\collaboration{RQCD Collaboration}
\noaffiliation

\begin{abstract}
We determine the leading order mesonic~($B_0$ and $F_0$) and baryonic~($m_0$, $D$ and $F$) SU(3) chiral perturbation theory low energy constants from lattice QCD. We employ gauge ensembles with $N_f=3$ (i.e., $m_u=m_d=m_s$) non-perturbatively improved Wilson fermions at six distinct values of the lattice spacing in the range $a\approx (0.039 - 0.098) \; \fm$, which constitute a subset of the Coordinated Lattice Simulations (CLS) gauge ensembles. The pseudoscalar meson mass $M_\pi$ ranges from around $430 \; \MeV$ down to $240 \; \MeV$ and the linear spatial lattice extent $L$ from $6.4\,M_{\pi}^{-1}$ to $3.3\,M_{\pi}^{-1}$, where $ L M_\pi \geq 4$ for the majority of the ensembles. This allows us to perform a controlled extrapolation of all the low energy constants to the chiral, infinite volume and continuum limits. We find the SU(3) chiral condensate and $F_0$ to be smaller than their SU(2) counterparts while the Gell-Mann--Oakes--Renner parameters $B_0\approx B$ are similar. Regarding baryonic LECs, we obtain $F/D = 0.612^{(14)}_{(12)}$.
\end{abstract}


\maketitle

\section{Introduction}

Chiral perturbation theory (ChPT) is a central tool for the description and understanding of a multitude of hadronic processes. In this context, the interplay between ChPT and lattice simulations of QCD is of particular interest: while in Nature the quark masses are fixed, in lattice simulations these (and other simulation parameters) can be varied and the precision and the range of validity of truncated ChPT expansions explored systematically. Moreover, some of the low energy constants (LECs) of this effective field theory can be constrained or determined from lattice data, which complements phenomenological fits to experimental data that are restricted to the physical quark mass point. Vice versa, ChPT augments lattice QCD simulations, providing parametrizations of the dependence of the results on the light quark masses and the simulation volume that are consistent with the dynamical breaking of chiral symmetry as well as with the global symmetries of QCD in the massless limit.

While the light pseudoscalar masses, decay constants, the chiral condensate and related mesonic quantities have been well explored in lattice QCD simulations and confronted with SU(2) ChPT predictions---see, e.g., the recent Flavour Lattice Averaging Group (FLAG) review~\cite{Aoki:2021kgd}---this is less so regarding baryonic observables. On the one hand, the lattice data are less precise for baryons, in particular towards small values of the quark masses. On the other hand, the number of independent LECs is larger and also the convergence properties of ChPT may be inferior in the baryonic sector. For instance, the mass gaps between octet and decuplet baryons are smaller than those between pseudoscalar mesons and vector meson resonances, which may necessitate the inclusion of decuplet baryons as explicit degrees of freedom, at least for some observables. Including hyperons, i.e., the $\Lambda$, the $\Sigma$ and the $\Xi$, into the ChPT analysis, in addition to the nucleon $N$ (or the $N$ and the $\Delta$ resonance), provides a wealth of additional information, whereas the number of baryonic LECs of flavour SU(3) ChPT increases only moderately relative to SU(2) ChPT. This makes SU(3) ChPT a particularly popular choice in the description of processes that involve  baryons. One concern regarding phenomenological applications, however, is the convergence of SU(3) ChPT at the physical point itself, where neither the mass $M_{\eta_8}\approx (\tfrac43 M_K^2-\tfrac13 M_\pi^2)^{1/2}\approx 565\;\MeV$ of the would-be $\eta_8$ pseudoscalar meson, the kaon mass $M_K\approx 494\;\MeV$ nor the average light meson mass $\overline{M}=(\tfrac23 M_K^2+\tfrac13 M_\pi^2)^{1/2}\approx 411\;\MeV$ are particularly small in comparison to the chiral symmetry breaking scale $\Lambda_\chi\coloneqq 4\pi F_0<4\pi F_{\pi}\approx 1160\;\MeV$. While this may limit the practical applicability of SU(3) ChPT regarding some observables, the corresponding LECs are well-defined and can in principle be obtained from lattice QCD.

Within most lattice simulations of $N_f=2+1$ (or of $N_f=2+1+1$) QCD the mass of the light quark $m_{\ell}=m_u=m_d$ is varied while that of the strange quark $m_s$ is kept approximately fixed near its physical value. In a few cases, instead $\tr M=m_u+m_d+m_s$ is kept constant~\cite{Bietenholz:2010jr,Bruno:2014jqa,Bruno:2016plf,Bickerton:2019nyz}. The former setting is ideal regarding SU(2) ChPT while neither choice is sufficient to determine SU(3) LECs, unless other quark mass combinations are added; in particular, one may want to reduce the trace of the mass matrix $\tr M$ below its physical value. This can be achieved via a partially quenched strategy, see, e.g., Refs.~\cite{WalkerLoud:2011ab,Beane:2011pc,liang2021detecting}, or, ideally, by realizing additional sea quark mass combinations~\cite{Bali:2016umi,Bali:2019svt}.

So far no comprehensive lattice QCD investigation of SU(3) ChPT exists, that includes pion masses smaller than $300\;\MeV$ or addresses the continuum limit---neither for mesons nor for baryons. Here we start to close this gap with a consistent, simultaneous analysis of several observables within the framework of SU(3) ChPT: we are in the process of computing the masses $M_{P}$ ($P\in\{\pi, K,\eta_8\}$) and $m_B$ ($B\in\{N, \Lambda, \Sigma, \Xi\}$) of the light pseudoscalar mesons and baryons as well as the corresponding decay constants $F_{P}$ and axial charges $g_A^B$ from $N_f=2+1$ QCD at many points in the plane spanned by the quark masses $m_{\ell}=m_u=m_d$ and $m_s$ at several values of the lattice spacing $a$.

Here we present first results, obtained on $N_f=3$ mass-degenerate gauge ensembles for the leading order (LO) mesonic LECs $F_0$ and $B_0$ and baryonic LECs $m_0$, $D$ and $F$, where $D$ and $F$ also enter the dependence of the octet baryon masses on the pseudoscalar meson masses at order $p^3$ (next-to-leading order (NLO) of heavy baryon ChPT (HBChPT) or next-to-next-to-leading order (NNLO) of covariant baryon ChPT (BChPT)). We remark that for $m_{\ell}=m_s$ all the octet baryons masses are degenerate, however, this is not so for the non-flavour singlet axial charges, where two independent combinations exist. The main quantity that determines the convergence properties of ChPT is the squared average pseudoscalar mass $\overline{M}\vphantom{M}^2$. The value realized in Nature corresponds to our largest quark mass values and we cover a range in $\overline{M}\vphantom{M}^2$ that extends down to less than one third of that: if SU(3) ChPT is applicable at the physical quark mass point then it should also apply to our lattice data, in the continuum limit.

The reliable determination of LO LECs from an extrapolation to the chiral limit requires at least NLO ChPT. Naturally, it is {\em a priori} unknown whether higher order ChPT may be required within the window of available pseudoscalar masses or if ChPT is applicable at all. Including higher orders is of limited practicability in view of the finite number of data points and their statistical errors, due to the exploding number of new LECs. However, simultaneously analysing a number of different quantities that should be sensitive to the same set of LECs like baryon masses and their axial charges can serve as a consistency check and reduces the parametric uncertainty. Here we attempt exactly this, albeit only for the LO LECs. Previous analyses of lattice QCD data that aimed at determining LECs focused on one type of observable at a time. Ideally, however, one would wish to confirm that the same set of LECs can be employed consistently across a range of quantities.

This article is organized as follows. In Sec.~\ref{sec:chiral} we collect all SU(3) ChPT expressions for the quark mass and volume dependence that are relevant for our analysis, restricting ourselves to the special case $m_{\ell}=m_s$. For completeness, additional expressions for the baryon mass and the axial charges are collected in Appendix~\ref{sec:axialadd}. Then, in Sec.~\ref{sec:lattice}, we discuss properties of the gauge ensembles employed, the analysis methods used, the non-perturbative renormalization and improvement of the pseudoscalar decay constant and the axial charges as well as our continuum and chiral limit extrapolation strategy. The determination of systematic errors through a model averaging procedure is detailed in Appendix~\ref{sec:error}. Finally, in Sec.~\ref{sec:results} we determine and discuss the LECs, before we conclude.

\section{Meson and baryon SU(3) ChPT expressions}
\label{sec:chiral}
\subsection{Infinite volume}
\label{sec:ifv}

Throughout this article the isospin limit $m_{\ell} = m_u = m_d$ is assumed and only the SU(3) symmetric case $m \coloneqq m_{\ell} = m_{s}$ is considered. Our aim is to determine the LO mesonic ($B_0$ and $F_0$) and baryonic ($m_0$, $D$ and $F$) SU(3) ChPT LECs. The ChPT expressions in which these LECs appear are conveniently expressed in terms of the quark mass-dependent variables
\begin{equation}
  x = \frac{2mB_0}{(4\pi F_0)^2} ,\quad
  \xi = \frac{M_\pi^2}{(4\pi F_0)^2} ,\quad
  \mathcal{L} = \log \left( \frac{M_\pi^2}{\mu^2} \right),\label{eq:abbrev}
\end{equation}
where $M_\pi$ denotes the pseudoscalar meson mass and $B_0 \coloneqq \Sigma_0/F_0^2$ the Gell-Mann--Oakes--Renner (GMOR) parameter, whereas \mbox{$\Sigma_0 \coloneqq -\left.\langle\bar{u}{u}\rangle\right|_{m=0}>0$} and \mbox{$F_0 \coloneqq \left. F_\pi\right|_{m=0}$} are the quark chiral condensate and the pseudoscalar decay constant, respectively, in the SU(3) chiral limit. The LO LECs do not depend on the scale $\mu$. For the analysis of the mesonic case, it is convenient to set $\mu^{-2}=8t_{0,{\rm ch}}$, using the Wilson scale parameter~$t_0$~\cite{Luscher:2010iy} in the chiral limit. From $t_{0,{\rm ch}}/t_0^*=1.037(5)$~\cite{inprep} and \mbox{$(8t_0^*)^{-1/2}=478(7)\;\MeV$}~\cite{Bruno:2017gxd}, where $t_0^\star$~\cite{Bruno:2016plf} is defined as the value of $t_0$ at the point where $12t_0^\star M_\pi^2 = 1.11$ (and $m_\ell = m_s$), we obtain $\mu=469(7)$~MeV. 

At NNLO in SU($N_f$) ChPT the corrections to the GMOR relation and the pion mass-dependence of the pseudoscalar decay constant~\cite{Gasser:1984gg,Amoros:1999dp,Bijnens:2013yca} read
\begin{align}
  M_\pi^2 &= 2B_0m
  [1 + x(a_{10} + a_{11}\mathcal{L})\nonumber\\ 
    &\qquad\qquad+ x^2(a_{20} + a_{21}\mathcal{L} + a_{22}\mathcal{L}^2)],
  \label{eq:gmor}\\
  F_\pi &= F_0 [ 
    1 + x(b_{10} + b_{11}\mathcal{L})\nonumber\\ 
    &\qquad\qquad+ x^2(b_{20} + b_{21}\mathcal{L} + b_{22}\mathcal{L}^2)],
  \label{eq:fp}
\end{align}
where
\begin{align}
  a_{11} &= \frac{1}{N_f}, & a_{22} &= \frac{9}{2N_f^2} - \frac{1}{2} + \frac{3N_f^2}{8}, \label{eq:fix1}
  \\
  b_{11} &= - \frac{N_f}{2}, & b_{22} &= - \frac{1}{2} - \frac{3N_f^2}{16}.\label{eq:fix2}
\end{align}
While $a_{10}$, $b_{10}$, $a_{21}$ and $b_{21}$ are combinations of NLO LECs, $a_{20}$ and $b_{20}$ are combinations of NNLO LECs. Whereas NLO and possibly NNLO corrections may turn out necessary to describe our lattice data for which $430\;\MeV\gtrsim M_\pi\gtrsim 240\;\MeV$, it needs to be seen whether all of these LECs can be resolved, in addition to lattice spacing effects.

The LO octet baryonic LECs are the nucleon mass in the chiral limit~$m_0$ and the couplings $F$ and $D$ which parameterize the octet axial charges in the SU(3) chiral limit and also enter within the chiral expansions of other octet baryon observables, in particular the masses. In the \mbox{$N_f=3$} flavour symmetric case at $\mathcal{O}(p^3)$ in BChPT the octet baryon mass~$m_B$ is given as~\cite{Ellis:1999jt,Lehnhart:2004vi}
\begin{align}
  m_B = m_0 + \bar{b} M_\pi^2 + 2\xi M_\pi\left(\frac{5D^2}{3}+3F^2\right)f_B\left(r\right) \label{eq:mb}
\end{align}
with $\bar{b}=-6b_0-4b_D$ being a combination of NLO LECs and $r=M_\pi/m_0$.
In the extended on-mass-shell (EOMS) scheme~\cite{Gegelia:1999gf,Fuchs:2003qc,Lehnhart:2004vi} the loop function is given as
\begin{align}
  f_B(r) = -2\left[\sqrt{1-\frac{r^2}{4}}\arccos\left(\frac{r}{2}\right)+\frac{r}{2}\log\left(r\right)\right],\label{eq:loopf}
\end{align}
where we follow the standard convention to identify the renormalization scale with $m_0$. Expanding this function for small $r$, i.e., for \mbox{$m_0\rightarrow\infty$}, one obtains the heavy baryon ChPT (HBChPT) limit~\cite{Gasser:1987rb,Bernard:1992qa} \mbox{$f_B(r)=-\pi+\mathcal{O}(r)$}. The EOMS BChPT expressions are also known at NNNLO~\cite{Ren:2012aj}, however, our present lattice data cannot constrain the additional free parameters.

Regarding the axial charges $g_A^B$, the pion mass dependence in the SU(3) case for the nucleon and the $\Sigma$ baryon at $\mathcal{O}(p^3)$ is given as~\cite{Jenkins:1990jv,Bijnens:1985kj,Ledwig:2014rfa}
\begin{align}
  g_A^N & = D + F + c_N\xi
    +\bar{c}_N \xi \log \left( \frac{M_\pi}{m_0} \right)
    + d_N \xi^{3/2}, \label{eq:d1}
    \\
  g_A^{\Sigma} &= 2F + c_{\Sigma}\xi 
    +\bar{c}_\Sigma \xi \log \left( \frac{M_\pi}{m_0} \right)
    + d_{\Sigma}\xi^{3/2}, \label{eq:d2}
\end{align}
where the coefficients,
{\small 
\begin{align}
  \bar{c}_N &= -\left[3(D+F)+\frac{1}{3}\left(27D^3+25D^2F+45DF^2+63F^3\right)\right] ,\label{eq:cn1} 
  \\
  \bar{c}_\Sigma &=-\left[6F+\frac{2}{3}F\left(25D^2+63F^2\right)\right],\label{eq:cn2}
\end{align}
}%
are entirely determined by the LO LECs. Above, $d_N=d_{\Sigma}=0$, however, such terms arise naturally when loop corrections that contain decuplet baryons are included~\cite{Jenkins:1991es}. For completeness, we reference the corresponding expectations (as well as those for $m_B$) in Appendix~\ref{sec:axialadd}. Unfortunately, these expressions, involving the additional LECs $\Delta$, $\mathcal{C}$ and $\mathcal{H}$, do not satisfactorily describe our data on $g_A^B$ while fits to $m_B$ suggest $\mathcal{C}\approx 0$. Including the logarithmic terms, a reasonable fit quality seems only possible when also adding the above phenomenological $d_B$-terms. However, such fits give very small values for $F$ and $D$, that are at variance with the pion mass-dependence of $m_B$. Leaving $\bar{c}_B$ as free parameters, i.e., ignoring the ChPT expectation, the data even suggest $\bar{c}_B>0$, opposite to the expectation of Eqs.~\eqref{eq:cn1} and~\eqref{eq:cn2}. Similar tensions are evident also in recent data on $g_A^N$ within SU(2) ChPT, see, e.g., Refs.~\cite{Chang:2018uxx,Gupta:2018qil,Lutz:2020dfi}. We interpret this as a sign of large cancellations between pion and decuplet loop effects, a full understanding of which requires to further reduce the quark mass and/or to increase the ChPT order. For the purpose of determining the LO LECs and also in view of the precision of the lattice data, we will truncate Eqs.~\eqref{eq:d1} and~\eqref{eq:d2} at $\mathcal{O}(p^2)$.

\subsection{Finite volume corrections}
\label{sec:fv}

Since ChPT also predicts the finite volume dependence, we include the associated corrections. For the pseudoscalar meson mass and decay constant in the continuum limit the dependence on the linear spatial lattice extent $L$ is given by~\cite{Gasser:1986vb,Gasser:1987zq}
\begin{align}
  M^2_\pi(L) &= M^2_\pi \left[ 1 + x \frac{1}{N_f} h(\lambda_\pi) + \cdots \right] , \label{eq:mpi_fv}
  \\
  F_\pi(L) &= F_\pi \left[ 1 - x \frac{N_f}{2} h(\lambda_\pi) + \cdots \right] \label{eq:fpi_fv}
\end{align}
with $M_\pi = M_\pi(L=\infty)$, $F_\pi = F_\pi(L=\infty)$ and to this order we can substitute $x$ for $\xi$. Above, $\lambda_\pi = LM_\pi$ and
\begin{align}
  h(\lambda) = 4 \sum_{\mathbf{n}\neq\mathbf{0}} 
  \frac{K_1(\lambda|\mathbf{n}|)}{\lambda|\mathbf{n}|}, \label{eq:h1}
\end{align}
where $\mathbf{n} \in \mathbb{Z}^3$ and $K_n(x)$ denotes the modified Bessel function of the second kind of order $n$. We will not consider two-loop finite volume effects~\cite{Colangelo:2005gd,Bijnens:2014dea} since these contain the NLO LECs.

For the octet baryon mass the SU(2) BChPT result~\cite{AliKhan:2003ack,Procura:2006bj} easily generalizes to SU(3):
\begin{align}
  m_B(L) =&\ m_B + 4 m_0 \xi
    \left( \frac{5D^2}{3} + 3F^2 \right) \nonumber \\
    &\cdot \int_0^{\infty}\!\!\!\dd{y} \sum_{\mathbf{n}\neq\mathbf{0}} K_0 
    \left(
      \lambda_\pi |\mathbf{n}|\sqrt{1-y+\frac{y^2}{r^2}}
    \right),
  \label{eq:mb_fv}
\end{align}
where we truncated the expression at $\mathcal{O}(p^3)$ and \mbox{$r=M_\pi/m_0$} as above. Note that corrections to the baryon mass $m_B$ due to transitions to decuplet baryons with the mass $m_{D0}$ were first considered in Ref.~\cite{Jenkins:1991es}. For completeness, we collect the corresponding $m_{\ell}=m_s$ expectations in Appendix~\ref{sec:axialadd}.

In the case of the axial charges $g_A^B$, the finite volume corrections given in Appendix~\ref{sec:ga_fv} have a sign opposite to the trend of the lattice data. It appears that---just like in the infinite volume case---the effect of decuplet baryons needs to be included, introducing three additional LECs which cannot be resolved at present. Therefore, we combine the infinite volume $\mathcal{O}(p^2)$ ChPT expectation with the dominant ChPT finite volume term
\begin{align}
  g_A^B(L) = g_A^B + c^{B}_V \xi \frac{\exp\left( -LM_\pi \right) }{\sqrt{LM_\pi}}, \label{eq:gA_fv}
\end{align} 
where $c^B_V$ is a free phenomenological coefficient.

\section{Lattice Set-Up}
\label{sec:lattice}

We discuss the gauge ensembles used. Subsequently, we summarize our determination of the relevant observables, including---where applicable---their renormalization and order $a$ improvement. We then list the results for the analysed ensembles and detail our continuum, infinite volume and chiral extrapolation strategy.

\subsection{Gauge ensembles}

In our analysis we employ ensembles generated with $N_f = 3$ flavours of non-perturbatively $\mathcal{O}(a)$-improved Wilson fermions with the tree-level Symanzik-improved gauge action. Most of the ensembles were produced within the Coordinated Lattice Simulations (CLS)~\cite{Bruno:2014jqa} effort. Here we only focus on the subset of ensembles with degenerate quark masses $m_u = m_d = m_s$.

The ensembles come with either periodic or open boundary conditions in time~\cite{Luscher:2011kk}, where the latter choice is necessary at the two finest lattice spacings to circumvent the freezing of the topological charge and thus to ensure ergodicity~\cite{Schaefer:2010hu}. On ensembles with open boundary conditions measurements are taken far away from the boundaries, where translational symmetry in time is restored within statistical precision.

\begin{figure}
\includegraphics[width=.5\textwidth]{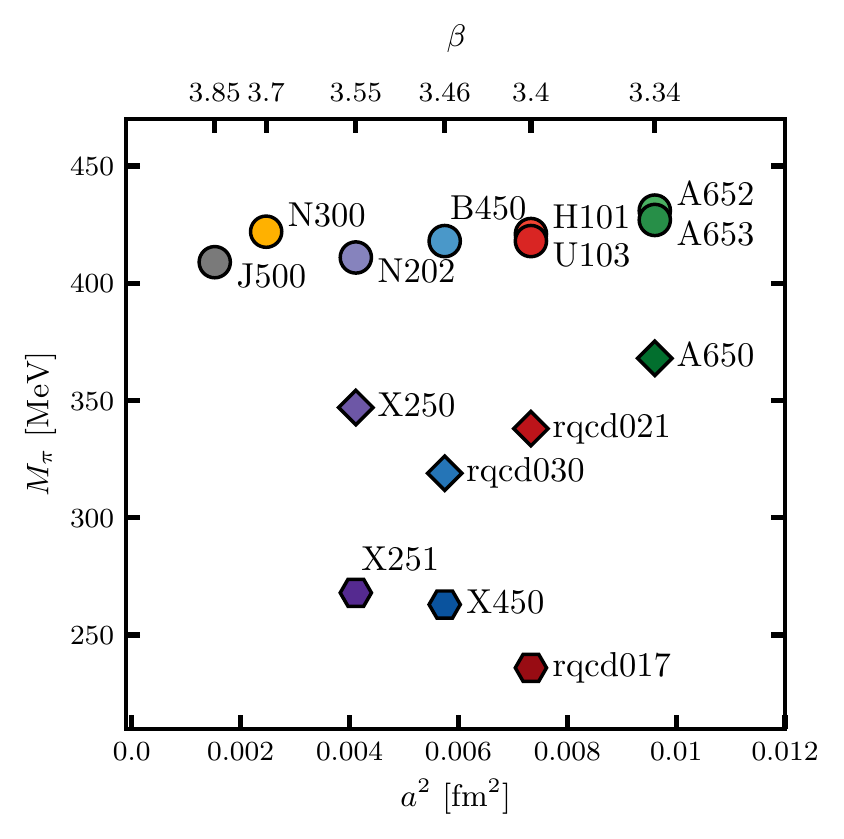}%
\caption{The parameter landscape of the ensembles listed in Table~\ref{tab:cls_ensembles}. The same colour coding will be used throughout this article to identify the individual ensembles. \label{fig:cls_ensemble_plot}}
\end{figure}

In total we analysed fifteen ensembles where the simulated parameter space is illustrated in Fig.~\ref{fig:cls_ensemble_plot}. More details can be found in Table~\ref{tab:cls_ensembles}. We cover a range of six different lattice spacings $0.039\;\fm\lesssim a\lesssim 0.098\;\fm$, the pion masses range from around $430 \; \MeV$ down to $240 \; \MeV$ and volumes are realized between $3.3 \leq L M_\pi  \leq 6.4$ where  $ L M_\pi \geq 4$ for the majority of the ensembles.
\begin{table}
\caption{The gauge ensembles analysed in this work. The rqcd{\tt xyz} ensembles were generated by RQCD using the {\sc BQCD} code~\cite{Nakamura:2010qh}, whereas all the other ensembles were generated within the CLS effort~\cite{Bruno:2014jqa}. The fourth column labels periodic (p) and open (o) boundary conditions, respectively. The lattice spacings $a$ were determined in Refs.~\cite{Bruno:2016plf,Bali:2016umi,inprep}.\label{tab:cls_ensembles}}
\begin{ruledtabular}
\begin{tabular}{lllcrcll}
Ensemble  & $\beta$ & $ a [\fm] $ & bc  & $N_t\cdot N_s^3$ & $M_\pi [\MeV]$ & $LM_\pi$ & $N_{\text{config}}$ \\
\hline
\mydot{A652}\!A652 & 3.34 & 0.098 & p & $48 \cdot 24^3$ & 431 & 5.14 & 4995 \\ 
\mydot{A653}\!A653 &      &    & p & $48 \cdot 24^3$ & 427 & 5.09 & 2525 \\ 
\mydiamond{A650}\!A650 &      &    & p & $48 \cdot 24^3$ & 368 & 4.4 & 2328 \\ 
\hline
\mydot{H101}\!H101 & 3.4 & 0.086 & o & $96 \cdot 32^3$ & 421 & 5.85 & 2000 \\ 
\mydot{U103}\!U103 &      &    & o & $128 \cdot 24^3$ & 418 & 4.35 & 2475 \\ 
\mydiamond{rqcd021}\!rqcd021 &      &    & p & $32 \cdot 32^3$ & 338 & 4.7 & 1541 \\ 
\myhexagon{rqcd017}\!rqcd017 &      &    & p & $32 \cdot 32^3$ & 236 & 3.27 & 2468 \\ 
\hline
\mydot{B450}\!B450 & 3.46 & 0.076 & p & $64 \cdot 32^3$ & 418 & 5.15 & 1612 \\ 
\mydiamond{rqcd030}\!rqcd030 &      &    & p & $64 \cdot 32^3$ & 319 & 3.94 & 1224 \\ 
\myhexagon{X450}\!X450 &      &    & p & $64 \cdot 48^3$ & 263 & 4.87 & 400 \\ 
\hline
\mydot{N202}\!N202 & 3.55 & 0.064 & o & $128 \cdot 48^3$ & 411 & 6.43 & 884 \\ 
\mydiamond{X250}\!X250 &      &    & p & $64 \cdot 48^3$ & 347 & 5.43 & 345 \\ 
\myhexagon{X251}\!X251 &      &    & p & $64 \cdot 48^3$ & 268 & 4.19 & 436 \\ 
\hline
\mydot{N300}\!N300 & 3.7 & 0.05 & o & $128 \cdot 48^3$ & 422 & 5.11 & 1520 \\ 
\hline
\mydot{J500}\!J500 & 3.85 & 0.039 & o & $192 \cdot 64^3$ & 409 & 5.2 & 751 
\end{tabular}
\end{ruledtabular}
\end{table}

\subsection{Analysis methods}
\begin{figure}
\includegraphics[width=.5\textwidth]{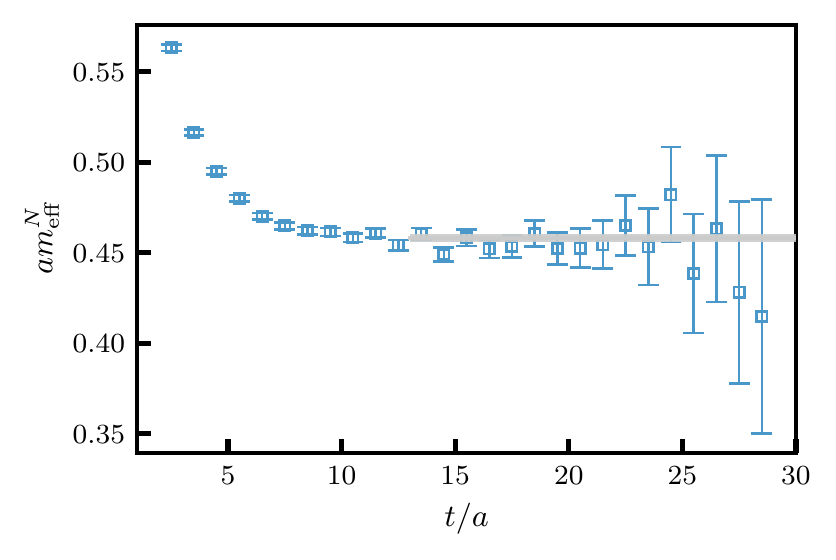}%
\caption{Effective mass (see Eq.~\eqref{eq:effmass}) of the baryon on ensemble B450. The grey horizontal error band indicates the fit range and the extracted ground state mass.
\label{fig:effmass}}
\end{figure}

The scale parameters $t_0/a^2$ and $t_0^\star/a^2$ as well as the quark mass from the axial Ward identity (AWI), the pseudoscalar meson mass and the octet baryon mass have been obtained within an extensive RQCD analysis~\cite{inprep} of the light hadron spectrum on all the available CLS gauge ensembles. For the present purpose we only require these results for the subset of $m_{\ell}=m_s$ ensembles. Details on the computation of the two-point correlation functions $C_{\text{2pt}}(t)$, the extraction of the ground state masses and the statistical methods applied to account for autocorrelation effects and to compute covariance matrices between these quantities will be described in Ref.~\cite{inprep}. In Fig.~\ref{fig:effmass} we show as an example the effective mass in lattice units for the nucleon
\begin{align}
  a m^N_{\text{eff}} (t + a/2) = \log\left( \frac{C_{\text{2pt}}(t)}{C_{\text{2pt}}(t+a)} \right),
  \label{eq:effmass}
\end{align}
together with the extracted ground state mass $am_N=am_B$, on the ensemble~B450. $C_{\text{2pt}}(t)$ in this case is a baryonic two-point function. For this, the pion two-point function and the baryon three-point functions, we employ Wuppertal smearing~\cite{Gusken:1989ad} at the source and the sink, using spatially APE-smeared~\cite{Falcioni:1984ei} gauge transporters. The root mean squared quark smearing radii range from about $0.6\;\fm$ (for $M_\pi\approx 420\;\MeV$) up to about $0.75\;\fm$ (for $M_\pi\approx 230\;\fm$), see Table~2 of Ref.~\cite{RQCD:2019jai}.

The pion decay constant and the AWI quark mass are also obtained from two-point functions, using similar methods as for the pseudoscalar mass. However, in this case the two-point functions are only smeared at the source. We follow the strategy detailed in Refs.~\cite{Bruno:2014lra} and~\cite{Collins:2017rhi}. The calculation of the octet axial charges~$g_A^B$ for the nucleon and the $\Sigma$ baryon is part of a long term project~\cite{Bali:2019svt}. The baryon three-point functions $C_{\text{3pt}}(t,\tau, J)$, are computed using the sequential source method~\cite{Maiani:1987by}, (approximately) realizing four distinct source-sink separations $t/\fm \in \{0.7, 0.8, 1.0, 1.2\}$ in order to control excited state contamination. The local current $J_{ud}= J_u-J_d$, where $J_q=\bar{q}\gamma_\mu\gamma_5 q$ is inserted at the time $\tau$. Note that since $m_u=m_d$, no quark line-disconnected contributions appear. For definiteness with respect to the quark content we choose $N=p\sim uud$, $\Sigma=\Sigma^+\sim uus$ and $\Xi=\Xi^0\sim ssu$. Since the Cartan subgroup of SU(3) has rank two, in the case of exact SU(3) flavour symmetry ($m_{\ell}=m_s$) all the axial charges $g_A^B$ can be written as combinations of just two fundamental charges $\overline{F}$ and $\overline{D}$:
\begin{align}
  g_A^{N}= \overline{F} +\overline{D},\quad g_A^{\Lambda}=0,\quad g_A^{\Sigma}=2\overline{F},\quad
  g_A^{\Xi}=\overline{F}-\overline{D}.
\end{align}
Here we choose $g_A^N$ and $g_A^{\Sigma}$ as our basis. The combinations
\begin{align}
  \overline{F}=\frac{1}{2}g_A^{\Sigma}\stackrel{m\rightarrow 0}{\longrightarrow}F,
  \quad
  \overline{D}=g_A^N-\frac{1}{2}g_A^{\Sigma}\stackrel{m\rightarrow 0}{\longrightarrow}D
\end{align}
approach the LECs $F$ and $D$ in the chiral limit.

\begin{figure}
\includegraphics[width=.5\textwidth]{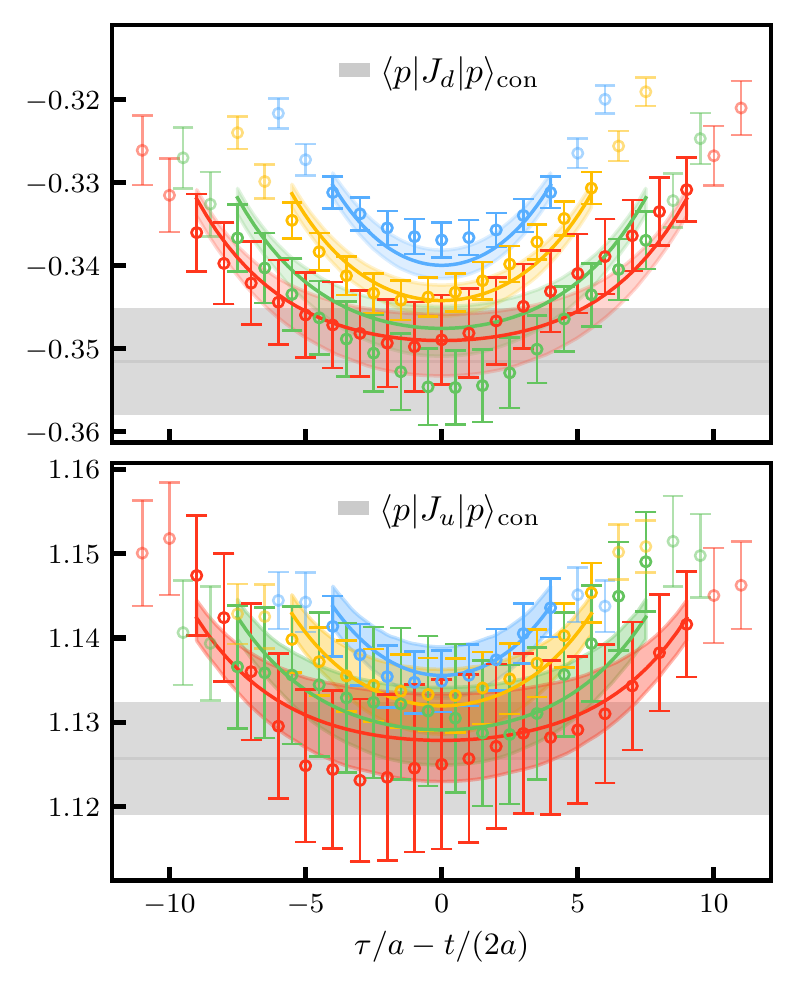}%
\caption{Simultaneous fit to all four source-sink separations of the ratios for $\langle p | J_d|p \rangle_{\text{con}} = \overline{F} - \overline{D}$ and $\langle p | J_u |p \rangle_{\text{con}} = 2\overline{F}$ on the ensemble N300. Only the dark symbols are included in the fit. The grey band shows the ground state contribution and its error.}
\label{fig:ratiofit}
\end{figure}

The matrix element of interest for a baryon $B$ can be obtained from a fit to the ratio of three-point over two-point functions
\begin{align}
    R^B(t,\tau, J_{ud}) 
    &= \frac{C^B_{\text{3pt}}(t,\tau,J_{ud})}{C^B_{\text{2pt}}(t)}\stackrel{t,\tau\rightarrow \infty}{\longrightarrow}g_A^B, \label{eq:rat}
\end{align}
see, e.g., Ref.~\cite{Bali:2014nma}, for details. As an example, we show in Fig.~\ref{fig:ratiofit} for the ensemble N300 a simultaneous fit for $J\in\{J_u, J_d\}$,\footnote{We take the differences of a proton with spin-up and spin-down along the direction $k$.} to the ratios
\begin{align}
    R^p_{\text{con}}(t,\tau, J) =
    b_{0,J}
    &+ b_{1,J} e^{-\Delta m\, t/2} \cosh\left( \Delta m ( \tau - t/2 ) \right) \nonumber \\
    &+ b_{2,J} e^{-\Delta m\, t}
\end{align}
for the proton, employing one and the same excited state mass gap $\Delta m$ in both channels, where the subscript ``con'' indicates that we only consider the quark line-connected Wick contractions. Exploiting the fact that all the quarks are mass-degenerate, this gives the matrix elements \mbox{$b_{0,J_u}=g_A^{\Sigma}=2\overline{F}$} and \mbox{$b_{0,J_d}=g_A^{\Sigma}-g_A^N=\overline{F}-\overline{D}$}.  The bootstrap error analysis is carried out using binned data with a bin size that is large compared to the integrated autocorrelation time, with the bootstraps matched to those of the other observables so that in the subsequent analysis all correlations can be taken into account.

\begin{table*}
\caption{
Results for the ensembles used in this work. The scale parameter $t_0/a^2$, the renormalized pion decay constant $F_\pi$ (where $F_\pi = f_\pi/\sqrt{2}$), the pion mass $M_\pi$, the baryon mass $m_B$, the RGI quark mass $m$ as well as the renormalized axial charges for the nucleon $g_A^N$ and the $\Sigma$ baryon $g_A^\Sigma$, respectively.
\label{tab:data}}
\begin{ruledtabular}
\begin{tabular}{llllllll}
Ensemble & $t_0 / a^2$ & $aF_\pi$ & $aM_\pi$ & $am_B$ & $am$ & $g_{A}^N$ & $g_{A}^\Sigma$ \\ 
\hline
A652    & 2.1697(56) & 0.04985(29) & 0.2140(10)  & 0.5842(41) & 0.02072(21)  &            &            \\ 
A653    & 2.1729(50) & 0.04980(25) & 0.21245(93) & 0.5855(37) & 0.02050(20)  & 1.1670(85) & 0.8903(61) \\ 
A650    & 2.2878(72) & 0.04598(36) & 0.1835(13)  & 0.5469(54) & 0.01547(21)  & 1.1489(94) & 0.8822(74) \\ 
H101    & 2.8545(81) & 0.04499(23) & 0.18286(57) & 0.5074(18) & 0.01796(10)  & 1.1818(87) & 0.9014(78) \\ 
U103    & 2.8815(57) & 0.04386(57) & 0.18158(60) & 0.5193(30) & 0.01745(10)  & 1.1334(74) & 0.8692(72) \\ 
rqcd021 & 3.032(15)  & 0.04084(23) & 0.14702(88) & 0.4508(47) & 0.01172(12)  & 1.1548(90) & 0.873(12)  \\ 
rqcd017 & 3.251(13)  & 0.03505(68) & 0.1022(15)  & 0.388(13)  & 0.00548(21)  &            &            \\ 
B450    & 3.663(11)  & 0.03999(13) & 0.16103(49) & 0.4582(24) & 0.016154(82) & 1.1723(58) & 0.8962(71) \\ 
rqcd030 & 3.914(15)  & 0.03535(18) & 0.12221(68) & 0.3957(90) & 0.009460(80) & 1.1437(89) & 0.8723(70) \\ 
X450    & 3.9935(92) & 0.03358(21) & 0.10144(62) & 0.3764(61) & 0.006574(57) & 1.175(10)  & 0.894(11)  \\ 
N202    & 5.165(14)  & 0.03419(18) & 0.13389(35) & 0.3799(18) & 0.013802(46) & 1.1806(58) & 0.9026(70) \\ 
X250    & 5.283(28)  & 0.03195(19) & 0.11321(39) & 0.3597(51) & 0.009880(47) & 1.1650(89) & 0.8884(93) \\ 
X251    & 5.483(26)  & 0.02932(21) & 0.08684(40) & 0.3185(85) & 0.005812(47) & 1.165(13)  & 0.889(14)  \\ 
N300    & 8.576(21)  & 0.02680(12) & 0.10647(38) & 0.3035(13) & 0.011332(30) & 1.1639(86) & 0.884(17)  \\ 
J500    & 14.013(40) & 0.02106(11) & 0.08119(34) & 0.2313(26) & 0.008755(21) & 1.1514(50) & 0.8873(84) \\ 
\end{tabular}
\end{ruledtabular}
\end{table*}

\subsection{Non-perturbative renormalization and improvement}

The quark mass, the pion decay constant and the axial charges need to be renormalized. We also $\mathcal{O}(a)$-improve these observables. Regarding the renormalization of the axial currents, we use the factors $Z_{A,sub}^l(g^2)$ of Ref.~\cite{DallaBrida:2018tpn}, obtained with the chirally rotated Schrödinger functional approach, as parameterized in their interpolation formula~(C.7). The renormalization factor $Z_M(g^2) = Z_A(g^2)/Z_P(g^2)$, required to translate the AWI quark mass $m^{\text{AWI}}$ into the renormalization group invariant (RGI)~\cite{Floratos:1978jb,Gasser:1982ap} mass~$m$, is given in Eq.~(5.6) of Ref.~\cite{Campos_2018}. We emphasize that both these factors have been computed entirely non-perturbatively. Using the improvement coefficients $b_A(g^2)$, $\tilde{b}_A(g^2)$, $b_P(g^2)$ and $\tilde{b}_P(g^2)$~\cite{Korcyl:2016cmx}, the observables can be renormalized and fully $\mathcal{O}(a)$-improved at each value of the lattice coupling $g^2=6/\beta$ as follows:
\begin{align}
  m &= Z_M \left[1 + a m^{\text{latt}} (b_A- b_P +3 \tilde{b}_A-3\tilde{b}_P)\right]m^{\mathrm{AWI}},\label{eq:im1}\\
  F_\pi &= Z_A \left[1 + a m^{\text{latt}} (b_A + 3 \tilde{b}_A) \right] F_\pi^{\text{latt}}, \\
  g_A^B &= Z_A \left[1 + a m^{\text{latt}} (b_A + 3 \tilde{b}_A) \right] g_A^{B,\text{latt}},\label{eq:im3}
\end{align}
where $am^{\text{latt}}=(\kappa^{-1}-\kappa_{\text{crit}}^{-1})/2$ is the lattice quark mass, $\kappa_{\text{crit}}$ is determined in Ref.~\cite{inprep} and we have assumed $m_{\ell}=m_s=\tfrac13 \tr M$. The uncertainties of the renormalization factors and improvement coefficients are incorporated in the statistical analysis by means of pseudo-bootstrap distributions.

\begin{table}
\caption{Values for $t_0^\star/a^2$ for each $\beta$-value taken from~\cite{inprep}.}
\label{tab:t0star}
\def\arraystretch{1.6}
\addtolength{\tabcolsep}{-1pt}
\begin{ruledtabular}
\begin{tabular}{ccccccc}
$\beta$ & 3.34 & 3.4 & 3.46 & 3.55 & 3.7 & 3.85 \\ \hline
  $\tfrac{t_0^\star}{a^2}$& 2.219(7) & 2.908(3) & 3.709(3) & 5.180(4) & 8.634(10) & 13.984(31) 
\end{tabular}
\end{ruledtabular}
\addtolength{\tabcolsep}{1pt}
\end{table}

\subsection{Lattice results\label{sec:lat_data}}

We will fit the squared pion mass $M^2_{\pi}$ and the pion decay constant $F_\pi$ simultaneously as functions of the RGI quark mass~$m$, whereas we parameterize the dependence of the baryon mass $m_B$ and of the axial charges $g_A^N$ and $g_A^\Sigma$ in terms of the pion mass. Regarding the continuum limit extrapolation, the quantities $t_0$ and $t_0^\star$ are required, as described below in more detail. In Table~\ref{tab:data} we summarize the corresponding results in lattice units for all the ensembles, with the exception of $t_0^\star/a^2$, listed in Table~\ref{tab:t0star}, whose values are common to all ensembles that share the same gauge coupling. Note that no axial charges have been determined on the ensembles A652 and rqcd017. However, ensemble A653 is very similar to A652 in terms of the simulation parameters while the rqcd017 volume is rather small and finite volume effects can be substantial, in particular for the axial charges.

\subsection{Extrapolation strategy}
\label{sec:extra}

A reliable extraction of the LO SU(3) LECs in the chiral limit requires a chiral, infinite volume and continuum limit extrapolation. Ideally, one would carry out simultaneous fits to all the observables. In particular, the mesonic LEC $F_0$ also appears within the ChPT expansions of the baryonic observables. In principle, this is possible and we even have the full covariance matrices available between $aM_\pi$,  $am$, $aF_\pi$, $am_B$, $g_A^N$ and $g_A^{\Sigma}$, however, the former three observables are much more precise in terms of their statistical accuracy than the baryonic ones. Therefore, any impact of the baryonic results onto the mesonic LECs should be negligible and we opt for a two stage procedure, first determining the mesonic LECs and then using the resulting value for $F_0/\sqrt{8t_{0,{\rm ch}}}$ within the extraction of the baryonic LECs.

For the action, the axial current (needed for $F_\pi$, $g_A^B$ and $m$) and the pseudoscalar current (needed for $m$), $\mathcal{O}(a)$~improvement is implemented non-perturbatively. Therefore, if we would simulate at a fixed lattice spacing~$a$, we would have full $\mathcal{O}(a)$~improvement. However, instead we keep the unimproved, bare lattice coupling $g^2$ fixed which results in a correction term $\propto  a\tr M$ for quantities~$aQ$, that are measured in lattice units.\footnote{In fact this mass-dependent shift of the improved lattice coupling also affects the renormalization factors of the axial and pseudoscalar currents but this effect has been accounted for within the definition of the improvement coefficients $\tilde{b}_A$ and $\tilde{b}_P$~\cite{Korcyl:2016cmx} of Eqs.~\eqref{eq:im1}--\eqref{eq:im3}.} This term cancels when constructing dimensionless combinations $(\sqrt{8t_0} a^{-1}) (aQ)$, using the scale parameter~$t_0/a^2$ on the same ensemble. Therefore, to achieve full $\mathcal{O}(a)$-improvement while varying the quark mass, we rescale all quantities $aQ \mapsto \sqrt{8t_0}Q$. This means that at the end of the analysis the dimensionful LECs $m_0$, $F_0$ and $B_0$ will be obtained in units of $\sqrt{8t_{0,{\rm ch}}}$, which can then be converted into physical units.

The continuum fit functions $X(\mathcal{M},L,a=0)$, where $\mathcal{M}=\sqrt{8t_0}m$ and $\mathcal{M}=8t_0 M_\pi^2$, respectively, for mesonic observables $X\in \{8t_0M_\pi^2, \sqrt{8t_0}F_\pi\}$ and baryonic observables $X\in\{\sqrt{8t_0}m_B, g_A^N, g_A^{\Sigma}\}$, are summarized in Eqs.~\eqref{eq:abbrev}--\eqref{eq:d2} and~\eqref{eq:mpi_fv}--\eqref{eq:gA_fv}. Note that the dependence \mbox{$t_0=t_{0,{\rm ch}}[1+k_1x+(k_{20}+k_{21}\mathcal{L})x^2+\ldots]$}~\cite{Bar:2013ora} does not interfere with the universal ChPT logs and therefore neither the functional forms of the continuum formulae nor the LECs are affected by the rescaling of all dimensionful quantities in units of $t_0$. Nevertheless, we remark that some of the higher order LECs, which we do not determine here, would require some knowledge about the LECs $k_1$ etc., that are associated with $t_0$. Regarding the lattice spacing-dependence, we assume the factorization
\begin{align}
  X(\mathcal{M},L,a)&=X(\mathcal{M}, L, 0)\label{eq:ansatz}\\\nonumber
  &\qquad \cdot \left[ 1 + \frac{a^2}{8t_0^\star}\left(c_a^X  +  \bar{c}_{a}^X 8t_0 M_\pi^2 \right)  \right]
\end{align}
into the continuum parametrization times mass-independent and mass-dependent lattice spacing effects, where $c_a^X$ and $\bar{c}_{a}^X$ are independent fit parameters for each observable $X$.

We will estimate the systematic errors of the LECs by varying the fit model and by employing different cuts on the ensembles that enter the fit:
\begin{enumerate}
  \item no cut: including all the available data points, 
  \item pion mass cut: excluding all ensembles with $M_\pi > 400\; \MeV$,
  \item lattice spacing cut: excluding the coarsest lattice spacing, i.e., the ensembles with $a\approx 0.098\;\fm$,
  \item volume cut: excluding all ensembles with $LM_\pi < 4$.
\end{enumerate}
We then carry out the model averaging procedure described in Appendix~\ref{sec:error}.

\section{Results and discussion}
\label{sec:results}

We determine the LO SU(3) mesonic LECs as well as the LO SU(3) octet baryonic LECs and compare the results with values from the literature.

\subsection{Mesonic LECs}
\label{sec:mesonic}
\begin{figure}
\includegraphics[width=.5\textwidth]{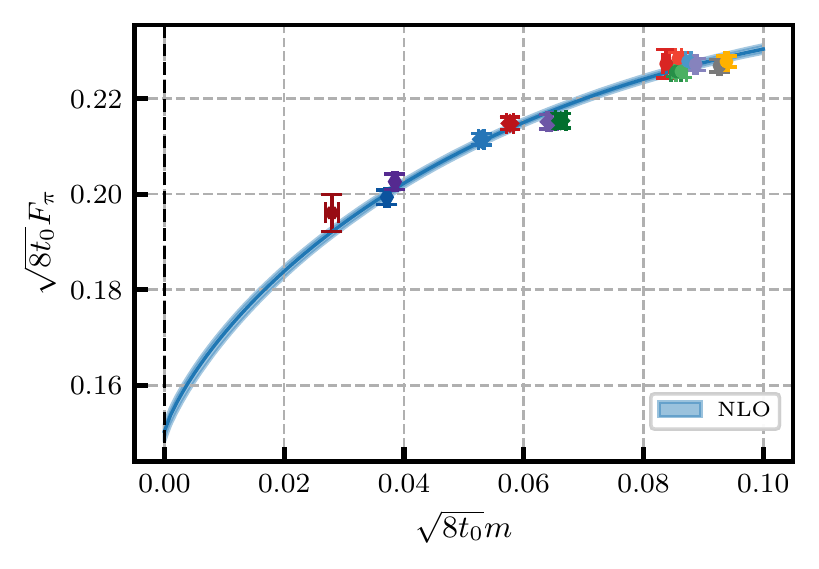}%
\caption{Extrapolation of the pion decay constant $F_\pi$ to the chiral limit. The data points are corrected for discretization and finite volume effects according to the parameters obtained from a combined fit to the pseudoscalar decay constant and mass on all the available data points employing the NLO ChPT ansatz. The blue band shows the NLO expression for the quark mass dependence.\label{fig:extrapolation_f0}}
\end{figure}
\begin{figure}
\includegraphics[width=.5\textwidth]{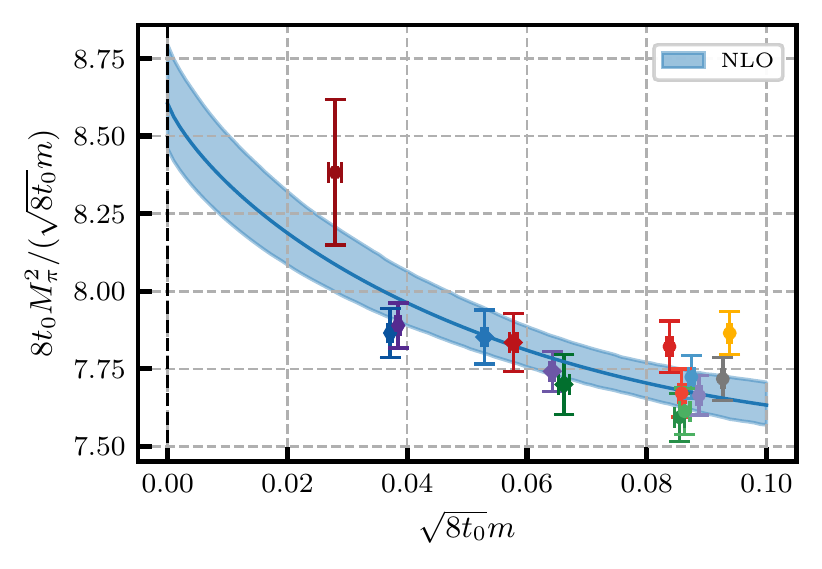}%
\caption{The same as Fig.~\ref{fig:extrapolation_f0} for the ratio of the squared pion mass over the quark mass~$M_\pi^2/m$.\label{fig:extrapolation_b0}}
\end{figure}
\begin{figure*}
\includegraphics[width=.56\textwidth]{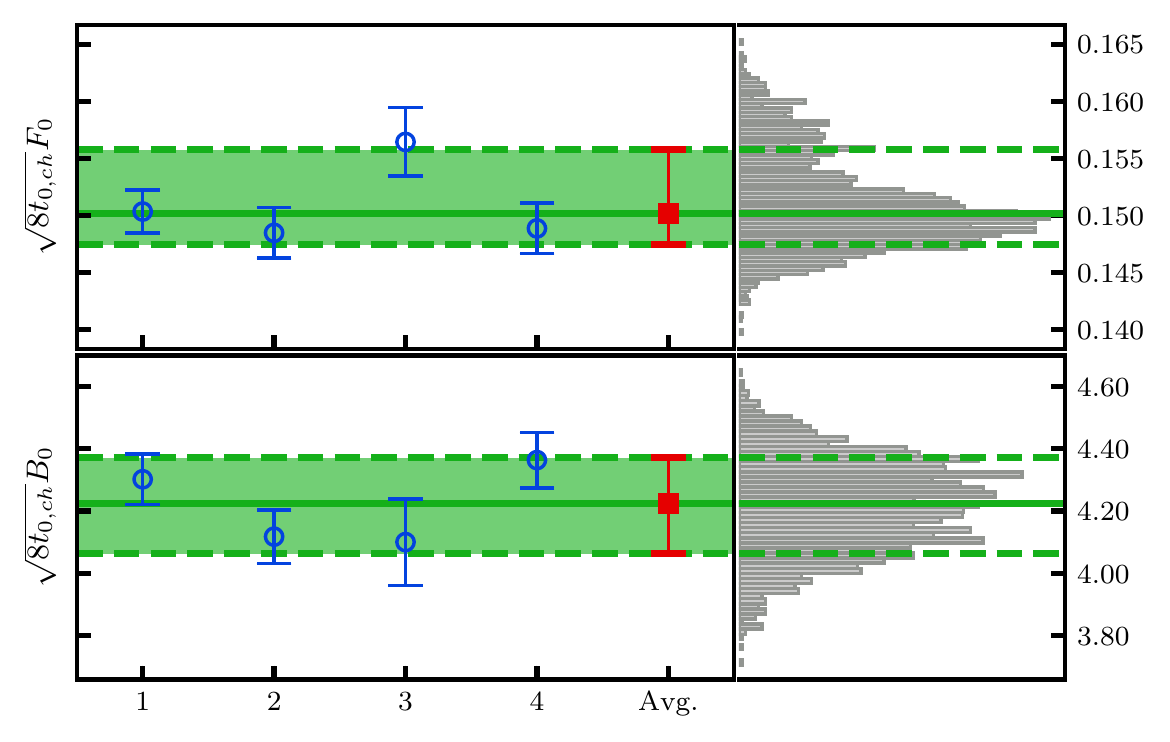}%
\caption{Final result for $F_0$ and $B_0$ (red point and green error band) obtained from individual fits (blue points) by performing the model averaging procedure described in Appendix~\ref{sec:error}. The model averaged distribution is shown as a histogram on the right where also the median and the 68\% confidence level interval are indicated (green lines).\label{fig:f0b0_weighted}}
\end{figure*}

The LO mesonic LECs $B_0$ and $F_0$ are determined by simultaneous fits to the pseudoscalar mass and decay constant as functions of the quark mass, the volume and the lattice spacing as described above. The fits are carried out including the errors of and the correlations between the pion decay constant, the pseudoscalar mass and the quark mass within each ensemble. The resulting $\chi^2$-values are fully correlated.

Including only the mass-independent discretization terms of Eq.~\eqref{eq:ansatz} and carrying out fits employing the NLO ChPT expressions, i.e., truncating the quark mass and the volume dependence at $\mathcal{O}(x)$, we are able to resolve all parameters reasonably well. Figures~\ref{fig:extrapolation_f0} and~\ref{fig:extrapolation_b0} illustrate the resulting quark mass-dependence of the pseudoscalar decay constant and the squared pion mass, respectively, from a combined fit to all the available data points. This fit to 30 points requires six parameters ($\sqrt{8t_0}B_0$, $\sqrt{8t_0}F_0$, $a_{10}$, $b_{10}$, $c_a^{M_\pi}$ and $c_a^{F_\pi}$) while the coefficients of the logs, $a_{11}=1/3$ and $b_{11}=-3/2$, are fixed, see Eqs.~\eqref{eq:fix1} and~\eqref{eq:fix2}. For a better visualization of the deviations from the linear GMOR, in Fig.~\ref{fig:extrapolation_b0} we have divided the squared pion mass by the quark mass (all in units of $8t_0$). This ratio approaches the GMOR expectation $2B_0\sqrt{8t_{0,{\rm ch}}}$ in the chiral limit. The deviation from a linear dependence is caused by $b_{11}$. This, as well as the curvature observed in Fig.~\ref{fig:extrapolation_f0} that is due to $a_{11}$, is in agreement with the data.

Since this simple fit describes the data very well, adding further parameters does not improve the situation: allowing for the mass-dependent discretization terms $\bar{c}_a^X\neq0$ in Eq.~\eqref{eq:ansatz}, does not significantly change the values of $\chi^2/N_{\text{dof}}$, $F_0$ or $B_0$. However, the errors for the fit parameters $c_A^X$, $a_{10}$ and $b_{10}$ increase considerably and on the reduced data sets, when incorporating the cuts described at the end of Sec.~\ref{sec:extra}, stable fits become impossible. Similarly, when allowing for the $\mathcal{O}(x^2)$ (NNLO) terms in the continuum fit functions~\eqref{eq:gmor} and~\eqref{eq:fp}, the statistical errors of all parameters increase while the higher order parameters are either comparable with zero or cannot be resolved reliably due to cancellations. After exploring these alternative parametrizations, we decided, in view of the range and quality of the present data, only to include the four parameter NLO continuum fit in conjunction with the two parameters that account for mass-independent $\mathcal{O}(a^2)$ effects into our analysis, and to explore the parametrization uncertainty by imposing the cuts on the data that are defined in Sec.~\ref{sec:extra}. Carrying out the fits on these four sets of ensembles and performing the model averaging procedure as described in Appendix~\ref{sec:error}, we obtain
\begin{align}
  \sqrt{8t_{0,{\rm ch}}} F_0 = 0.1502^{(56)}_{(29)},\quad
  \sqrt{8t_{0,{\rm ch}}} B_0 = 4.22^{(15)}_{(16)},
\end{align}
where the errors include the systematics. The individual results for each fit are listed in Table~\ref{tab:fit_results_NLO_meson} and compiled in Fig.~\ref{fig:f0b0_weighted}, where also the final result is indicated.
\begin{table}
\caption{Results for the LO mesonic LECs $F_0$ and $B_0$ in units of $1/\sqrt{8t_{0,{\rm ch}}}=469(7)\,$MeV obtained from fits to the NLO ChPT expression and different subsets of the parameter space spanned. The subsets are defined at the end of Sec.~\ref{sec:extra}.\label{tab:fit_results_NLO_meson}}
\def\arraystretch{1.15}
\begin{ruledtabular}
\begin{tabular}{clll}
Fit & $\chi^2/N_{\text{dof}}$ & $\sqrt{8t_{0,{\rm ch}}} F_0$ & $\sqrt{8t_{0,{\rm ch}}} B_0$  \\
\hline
1 & 0.9322 & $ 0.1504(19)  $ & $ 4.302(81) $ \\
2 & 0.7146 & $ 0.1565(30)  $ & $ 4.10(14) $ \\
3 & 0.3444 & $ 0.1485(22)  $ & $ 4.118(86) $ \\
4 & 1.0500 & $ 0.1489(22)  $ & $ 4.364(89) $ \\
\end{tabular}
\end{ruledtabular}
\end{table}

\subsection{Baryonic LECs}

\begin{figure}
\includegraphics[width=.5\textwidth]{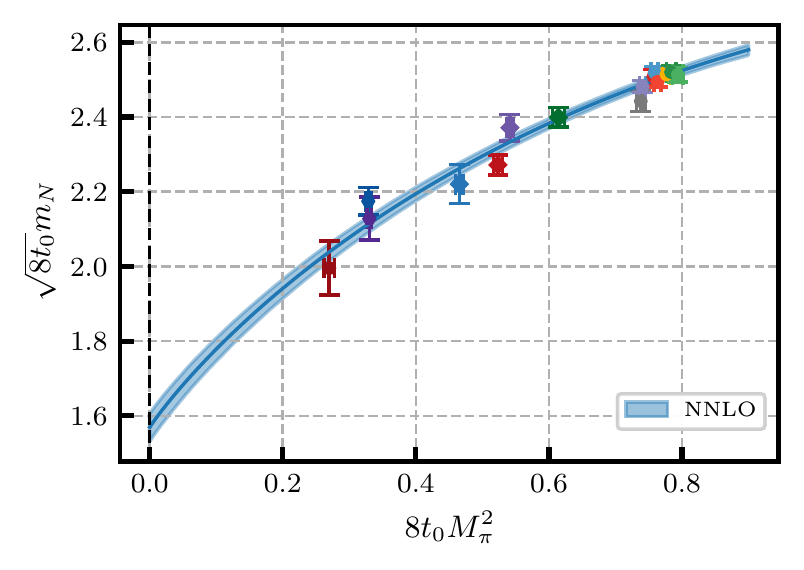}%
\caption{Extrapolation of the nucleon mass $m_N$ to the chiral limit. The data points are corrected for discretization and finite volume effects according to the parameters obtained from a combined fit to the nucleon mass and the two axial charges on all the available data points. The blue band shows the NNLO BChPT expression for the pion mass dependence.\label{fig:extrapolation_m0}}
\end{figure}

\begin{figure}
\includegraphics[width=.5\textwidth]{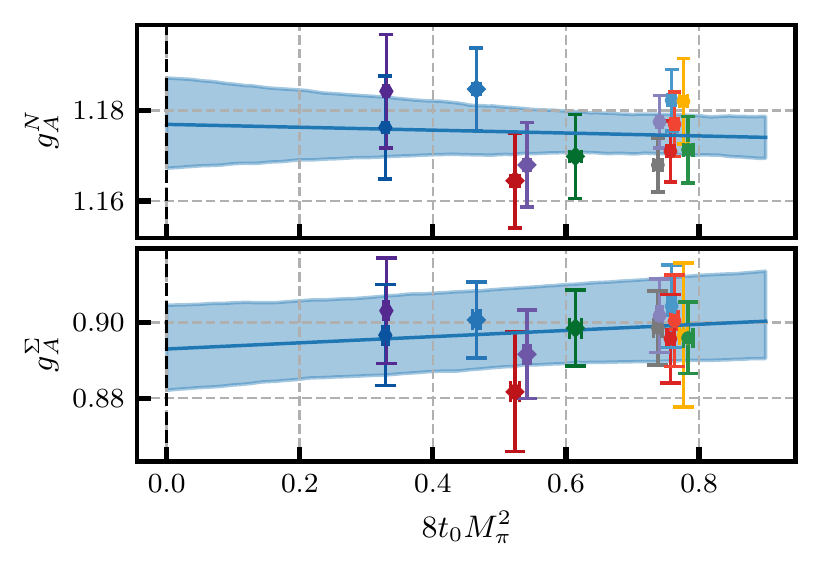}%
\caption{The same as Fig.~\ref{fig:extrapolation_m0} for the axial charges of the nucleon and the $\Sigma$ baryon. The blue band shows the NLO ($\mathcal{O}(p^2)$) chiral extrapolation.\label{fig:extrapolation_ga}}
\end{figure}

In analogy to the analysis of the mesonic observables, we carry out a simultaneous extrapolation of the octet baryon mass and the axial charges for the nucleon and the $\Sigma$ baryon. The continuum expressions for the dependence of these three observables on the pion mass and the lattice extent $L$ are given in Eqs.~\eqref{eq:mb}--\eqref{eq:d2},~\eqref{eq:mb_fv} and~\eqref{eq:gA_fv}. Again, lattice spacing effects are parameterized as in Eq.~\eqref{eq:ansatz}. For the decay constant $F_0$, that enters in the definition of $\xi$, we use the result obtained in Sec.~\ref{sec:mesonic}. HBChPT should give the same set of LO LECs $m_0$, $F$ and $D$ as BChPT in the EOMS prescription. To investigate the impact of different truncations of the chiral expansion, in addition to the BChPT fits, we also carry out a HBChPT analysis, replacing the loop function~\eqref{eq:loopf} $f_B(r)\mapsto -\pi$.

The pion mass dependence of the axial charges appears to be mild. As already pointed out at the end of Sec.~\ref{sec:ifv}, the logarithmic corrections suggested by ChPT without decuplet loops differ in sign from what the data suggest and this---within our window of pion masses---can only be compensated for by corrections of $\mathcal{O}(\xi^{3/2})$ and higher and/or by including effects of the decuplet baryons, adding the additional LECs $\Delta$, $\mathcal{C}$ and $\mathcal{H}$. The same observation is made regarding finite volume effects, whose sign can only be reconciled with the data if decuplet loops are included. We list the relevant formulae in Appendix~\ref{sec:ga_fv} but we cannot explore these additional contributions, given the statistical error of our present data. Therefore, regarding the axial charges, we opt for the NLO ($\mathcal{O}(p^2)$) analysis and truncate Eqs.~\eqref{eq:d1} and~\eqref{eq:d2} at $\mathcal{O}(\xi)$. Regarding the finite volume effects, we restrict ourselves to the leading term~\eqref{eq:gA_fv}, with phenomenological coefficients $c_V^N$ and $c_V^{\Sigma}$. Turning to the baryon mass, we are able to employ the full NNLO ($\mathcal{O}(p^3)$) expressions, both for the pion mass-dependence and the finite volume behaviour. We also found the baryon mass data to be well described when including decuplet loops, however, in this case, the LEC $\mathcal{C}$ is found to be compatible with zero within large errors, suggesting that the impact of the decuplet on the octet baryon mass is small.

\begin{figure*}
\includegraphics[width=.56\textwidth]{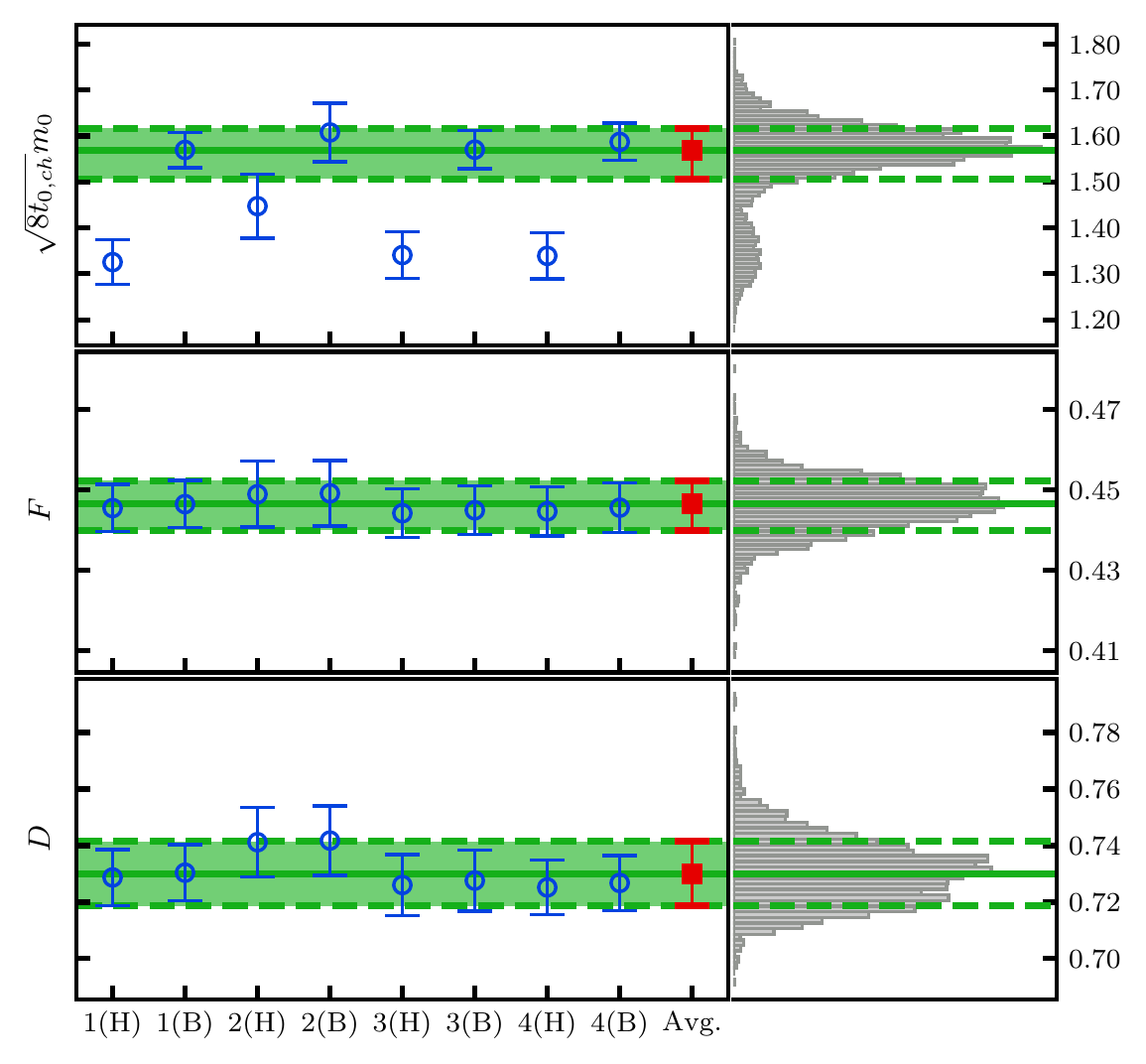}%
\caption{The same as Fig.~\ref{fig:f0b0_weighted} but for $m_0$, $F$ and $D$. For each cut there are two data points: BChPT~(B) and HBChPT~(H).\label{fig:m0ga_weighted}}
\end{figure*}

In Figs.~\ref{fig:extrapolation_m0} and~\ref{fig:extrapolation_ga} the pion mass dependencies of the nucleon mass and of the axial charges are shown, respectively, for a combined fit to all the available data points. The fit is to 41 data points (15 ensembles for $m_B$ and 13 ensembles for each of the axial charges) and requires 11 parameters, $m_0$, $F$, $D$, $\bar{b}$, $c_N$, $c_{\Sigma}$, $c_V^N$, $c_V^{\Sigma}$, $c_a^N$, $c_a^{g_A^N}$ and $c_a^{g_A^{\Sigma}}$: six (combinations of) LECs, two finite volume parameters for the axial charges and three parameters to describe discretization effects. We carry out the same variations of the data set as in the meson case. In addition, we explore both BChPT and HBChPT for the pion mass-dependence of the baryon mass, giving eight distinct results that are collected in Table~\ref{tab:fit_results_baryon} and shown in Fig.~\ref{fig:m0ga_weighted}. We find BChPT to give better fit qualities than HBChPT which is why the former fits dominate the averaging procedure. The BChPT results for $m_0$ are systematically larger than those of HBChPT which suggests a larger curvature of the data. Since $D$ and $F$ are mostly determined by the axial charges, where to the order that we employ no difference between BChPT and HBChPT exists, these values are largely unaffected by the parametrization. The final, averaged results read:
\begin{align}
    \sqrt{8t_{0,{\rm ch}}} m_0 = 1.57^{(5)}_{(6)}, \quad
    F = 0.447^{(6)}_{(7)}, \quad
    D = 0.730^{(11)}_{(11)}.\label{eq:FD}
\end{align}
Again, the errors include the systematics of the extrapolation.

\begin{table}
\caption{Results for the LO baryonic LECs $m_0$ (octet baryon mass in the chiral limit), $F$ and $D$ obtained from fits to the BChPT~(B) and HBChPT~(H) expressions on different subsets of ensembles. The subsets are defined at the end of Sec.~\ref{sec:extra}.\label{tab:fit_results_baryon}}
\def\arraystretch{1.15}
\begin{ruledtabular}
\begin{tabular}{lllll}
Fit & $\chi^2/N_{\text{dof}}$ & $\sqrt{8t_{0,{\rm ch}}} m_0$ & $F$  & $D$ \\
\hline
1 (H) & 1.1710 & $ 1.325(49)  $ & $ 0.4455(59) $ & $ 0.729(10) $\\
1 (B) & 0.9451 & $ 1.570(39)  $ & $ 0.4465(59) $ & $ 0.730(10) $\\
2 (H) & 1.4793 & $ 1.447(70)  $ & $ 0.4489(82) $ & $ 0.741(12) $\\
2 (B) & 1.2450 & $ 1.608(64)  $ & $ 0.4492(82) $ & $ 0.742(12) $\\
3 (H) & 1.3788 & $ 1.341(51)  $ & $ 0.4442(61) $ & $ 0.726(11) $\\
3 (B) & 1.1174 & $ 1.570(42)  $ & $ 0.4449(61) $ & $ 0.728(11) $\\
4 (H) & 1.2265 & $ 1.339(50)  $ & $ 0.4447(61) $ & $ 0.725(10) $\\
4 (B) & 0.9689 & $ 1.587(41)  $ & $ 0.4456(61) $ & $ 0.727(10) $\\
\end{tabular}
\end{ruledtabular}
\end{table}

\subsection{Comparison with other recent determinations}

\begin{figure}
\includegraphics[width=.5\textwidth]{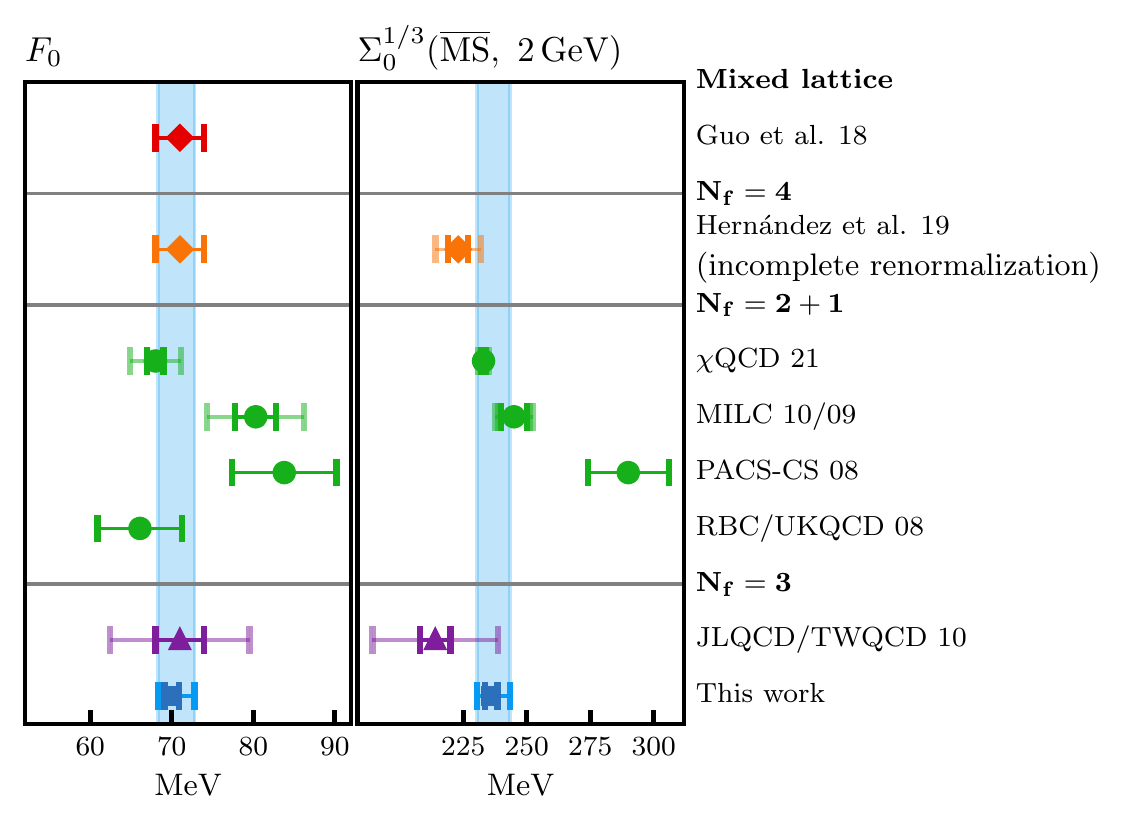}%
\caption{Comparison with the most recent SU(3) ChPT determinations of $F_0$ and $\Sigma_0 = B_0F_0^2$ from other groups. The latter is in the $\overline{\text{MS}}$ scheme at the scale $2\;\GeV$ with three active flavours. Note that the result labelled ``$N_f=4$'' is for the $N_f=3$ LECs, however, extrapolated from $N_f=4$ simulations at different numbers of colours. Dark error bars correspond to the statistical error only, whereas the lighter error bars include a systematic error estimate, added in quadrature.\label{fig:compare_f0b0_groups}}
\end{figure}

We employ the value $(8t_{0,{\rm ch}})^{-1/2}=\mu=469(7)\;\MeV$ to convert our results into physical units. As explained in Sec.~\ref{sec:ifv}, this value is obtained by combining $t_{0,{\rm ch}}/t_0^*=1.037(5)$~\cite{inprep} with \mbox{$(8t_0^*)^{-1/2}=478(7)\;\MeV$}~\cite{Bruno:2017gxd}. The mesonic LECs (with systematic uncertainties included in the errors) then read
\begin{align}
  F_0 = 70^{(3)}_{(2)} \; \MeV,\quad\Sigma_0^{1/3}(\text{RGI}) = 214^{(7)}_{(5)} \; \MeV,
\end{align}
where $\Sigma_0 = B_0F_0^2$. Note that $\Sigma_0(\text{RGI})$ refers to the value of the chiral condensate in the RGI scheme with $N_f=3$ active sea quark flavours. Using version~3 of the {\sc Mathematica} implementation of the {\sc RunDec} package~\cite{Herren:2017osy,Chetyrkin:2000yt} at five loop accuracy in the quark mass anomalous dimension- and the $\beta$-functions, we obtain the conversion factor $m(\text{RGI})=1.330(14)(7)m(\overline{\text{MS}}, 2\,\GeV)$ for the quark mass between the RGI and the $\overline{\text{MS}}$ schemes.\footnote{The normalization of the RGI mass used in {\sc RunDec3} differs from the one we employ. References~\cite{Floratos:1978jb,Gasser:1982ap,Aoki:2021kgd} share our convention.} The first error corresponds to the uncertainty of the three-flavour $\Lambda$-parameter~\cite{Bruno:2017gxd}, whereas the second error is the difference between five- and four-loop running. Using the scale-independence of $m\Sigma_0$ and taking the third root, we obtain
\begin{align}
  \Sigma_0^{1/3}(\overline{\text{MS}}, 2 \, \GeV)   &= 236^{(7)}_{(6)} \; \MeV .
\end{align}
Fig.~\ref{fig:compare_f0b0_groups} shows a comparison of our results for $F_0$ and $\Sigma_0$ with the most recent determinations from SU(3) ChPT analyses of other groups, also see the present FLAG~report~\cite{Aoki:2021kgd} for a detailed discussion. One issue with $N_f=2+1(+1)$ simulations is that the strange quark mass is usually kept close to its physical value, which limits the sensitivity of observables to the deviation of $F_0$ and $B_0$ from their SU(2) ChPT counter parts and necessitates partially quenched analyses. The only other simulation with $N_f=3$ mass-degenerate quarks was carried out over a decade ago by JLQCD/TWQCD~\cite{Fukaya:2010na}.

From an analysis of several lattice data sets Guo \textit{et}~al.~\cite{Guo:2018zvl} estimated $F_0 = 71(3)\; \MeV$. Hern\'{a}ndez \textit{et}~al.~\cite{Hern_ndez_2019} find from a large $N_c$ scaling analysis of $N_f=4$ and $N_c=3$--$6$ lattice data $F_0 = 71(3) \;\MeV$ and $\Sigma_0^{1/3} = 223(4)(8) \;\MeV$ for $N_f = N_c=3$. Simulating $N_f = 3$ flavours, JLQCD/TWCQD~\cite{Fukaya:2010na} determine $F_0 = 71(3)(8) \;\MeV$ and $\Sigma_0 = 214(6)(24) \;\MeV$. Employing $N_f = 2 + 1$ flavour simulations, the most recent determinations of $F_0$ are $68(1)(3) \;\MeV$ by $\chi$QCD~\cite{liang2021detecting}, $80.3(2.5)(5.4) \;\MeV$ by MILC~\cite{milccollaboration2010results}, $66.1(5.2) \;\MeV$  by RBC/UKCQD~\cite{Allton_2008} and $83.8(6.4) \;\MeV$  by PACS-CS~\cite{Aoki_2009}. For $\Sigma_0^{1/3}$ in the $\overline{\text{MS}}$ scheme at $2\;\GeV$, $\chi$QCD~\cite{liang2021detecting} find $233(1)(2) \;\MeV$, MILC~\cite{bazavov2009milc} quote $245(5)(4)(4) \;\MeV$, while PACS-CS~\cite{Aoki_2009} report $290(16) \;\MeV$. In summary, all the results for the mesonic LECs agree within their errors, with the exception of PACS-CS~\cite{Aoki_2009}, in particular regarding the chiral condensate.

\begin{figure}
\includegraphics[]{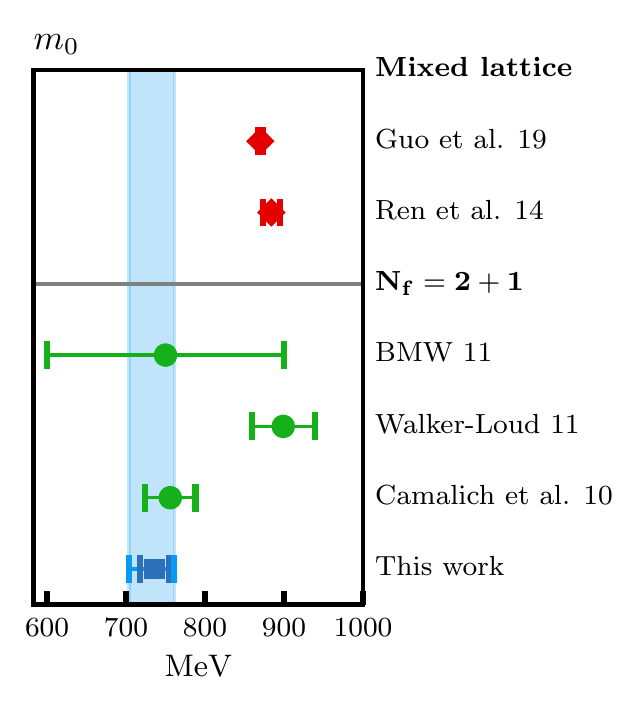}%
\caption{Comparison with the most recent determinations of the octet baryon mass in the $N_f=3$ chiral limit $m_0$, obtained from fits to Lattice QCD results.\label{fig:compare_m0_groups}}
\end{figure}

A compilation of the most recent results for the octet baryon mass in the SU(3) chiral limit is shown in Fig.~\ref{fig:compare_m0_groups}. Our result, including the systematic uncertainties and converted into physical units, reads
\begin{align}
  m_0 &= 736^{(25)}_{(32)} \; \MeV .
\end{align}
Carrying out SU(3) HBChPT or BChPT analyses of data from $N_f=2+1$ flavour simulations for $m_0$, Walker-Loud~\cite{WalkerLoud:2011ab} predicts $899(40) \;\MeV$, BMW~\cite{Durr:2011mp} find $750(150) \;\MeV$ and Martin Camalich~\textit{et}~al.~\cite{MartinCamalich:2010fp} obtain $756(32) \;\MeV$. Investigating multiple lattice data sets, Guo~\textit{et}~al.~\cite{Lutz:2018cqo,Guo:2019nyp} obtain $870(3) \;\MeV$ (mean and error estimated from the two fit results quoted in Ref.~\cite{Guo:2019nyp}) and Ren~\textit{et}~al.~\cite{Ren:2014vea} $884(11) \;\MeV$. A number of earlier results exists~\cite{WalkerLoud:2008bp,Ishikawa:2009vc,Ren:2012aj,Ren:2013wxa,Lutz:2014oxa}, which are not displayed in the figure. While it is difficult to estimate realistic errors for the two very global fits to lattice data~\cite{Lutz:2018cqo,Ren:2014vea}, there is disagreement between our results and Walker-Loud~\cite{WalkerLoud:2011ab} who obtains a much larger value.

\begin{figure}
\includegraphics[width=.5\textwidth]{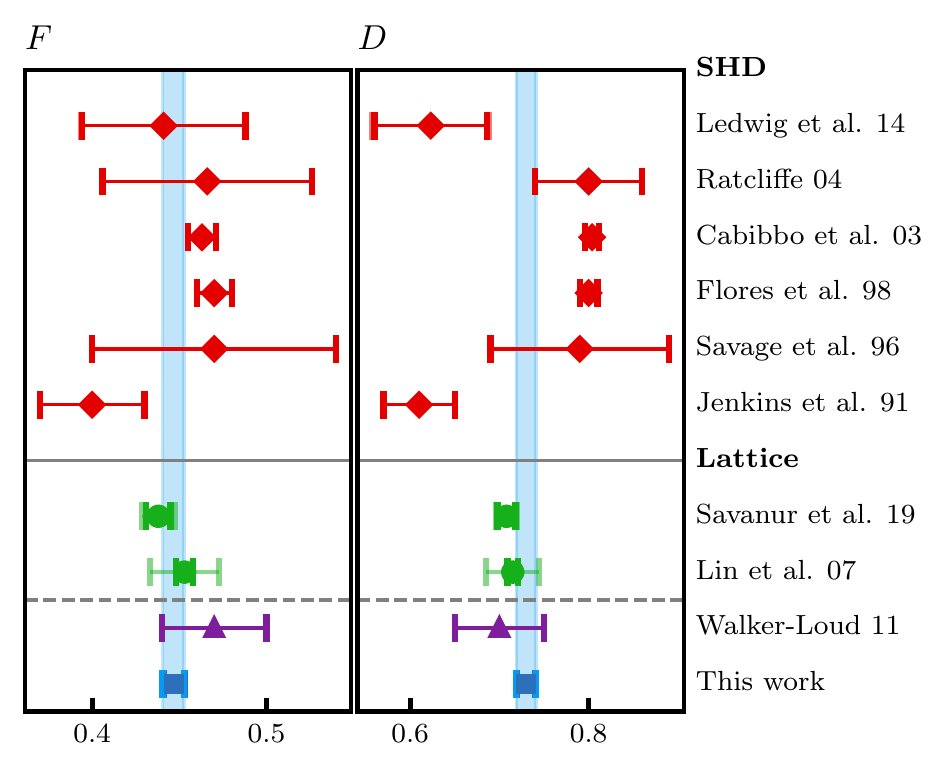}%
\caption{Comparison of our results for the LECs $F$ and $D$ with results obtained from lattice QCD calculations of the hyperon axial charges (green points)---albeit for physical quark masses, rather than in the chiral limit---and the baryon mass (purple point). In addition, we show selected results obtained from measurements of semileptonic hyperon decays.\label{fig:compare_ga_chiral_groups}}
\end{figure}

In Fig.~\ref{fig:compare_ga_chiral_groups} we compare our results~\eqref{eq:FD} for the baryonic LECs $F$ and $D$ with results obtained from lattice as well as phenomenological determinations. From a lattice QCD calculation of the axial charges, Lin and Orginos~\cite{Lin:2007ap} determine $F = 0.453(5)(19)$ and $D = 0.715(6)(29)$ with $N_f=2+1$ flavours. Later Savanur and Lin~\cite{Savanur:2018jrb} find $F = 0.438(7)(6)$ and $D = 0.708(1)(6)$, this time with $N_f=2+1+1$ flavours. Both values, however, refer to the physical quark mass point, where the definition of $F$ and $D$ is ambiguous, rather than to the chiral limit. From the baryon masses, Walker-Loud~\cite{WalkerLoud:2011ab} finds $F = 0.47(3)$ and $D = 0.70(5)$. Most phenomenological predictions are inferred from semileptonic hyperon decays. A selection of such analyses contains Jenkins~\textit{et}~al.~\cite{Jenkins:1991es}, Savage~\textit{et}~al.~\cite{Savage:1996zd}, Flores~\textit{et}~al.~\cite{FloresMendieta:1998ii}, Cabibbo~\textit{et}~al.~\cite{Cabibbo:2003cu}, Ratcliffe~\cite{Ratcliffe:2004jt} and Ledwig~\textit{et}~al.~\cite{Ledwig:2014rfa}. Regarding $F$, there is no clear contradiction when comparing any pair of results within the stated errors. With respect to $D$, however, Flores~\textit{et}~al.~\cite{FloresMendieta:1998ii} and Cabibbo~\textit{et}~al.~\cite{Cabibbo:2003cu}---while obtaining central values very similar to those of Savage~\textit{et}~al.~\cite{Savage:1996zd} and Ratcliffe~\cite{Ratcliffe:2004jt}---are at variance with the lattice determinations, within their errors. Note that the lattice results agree with each other, however, this should change if the precision was increased since two of the studies give numbers that correspond to the physical strange quark mass, rather than to the $N_f=3$ chiral limit.

\section{Summary and Outlook}

We carried out a simultaneous determination of all LO mesonic ($B_0$, $F_0$) and octet baryonic ($m_0$, $D$, $F$) SU(3) ChPT LECs, using $N_f = 3$ lattice QCD simulations. The analysis is based on fifteen gauge ensembles, spanning a range of pion masses from 430 MeV down to 240 MeV across six different lattice spacings between $a\approx 0.039\;\fm$ and $a\approx 0.098\;\fm$ and spatial lattice sizes between $3.3 \leq L M_\pi  \leq 6.4$. We found that a consistent description of the pion mass and volume dependence of the axial charges and the octet baryon mass was possible with the same set of LECs. Systematic errors were assessed and included by imposing cuts on the pion mass, the lattice spacing and the volume. For the baryon mass both covariant BChPT and HBChPT were employed. The resulting LECs are as follows ($\Sigma_0=F_0^2B_0$):
\begin{align*}
    &F_0 = 70^{(3)}_{(2)} \; \MeV,  \\
    &\Sigma_0^{1/3}  = 214^{(7)}_{(5)} \; \MeV \  (\text{RGI}), \\
    &\phantom{\Sigma_0^{1/3}}  = 236^{(7)}_{(6)} \; \MeV \ (\overline{\text{MS}}, 2 \; \GeV), \\
    &B_0=1.98^{(7)}_{(8)}\;\GeV\  (\text{RGI}),\\
    &\phantom{B_0}=2.63^{(10)}_{(10)}\;\GeV\ (\overline{\text{MS}}, 2 \; \GeV),
\end{align*}
\begin{align*}
    &m_0 = 736^{(25)}_{(32)} \; \MeV, \\ 
    &F = 0.447^{(6)}_{(7)}, \\ 
    &D = 0.730^{(11)}_{(11)},\\
    &\frac{F}{D} = 0.612^{(14)}_{(12)},
\end{align*}
where the uncertainties of the continuum, chiral and infinite volume extrapolation as well as of the conversion into physical units are included in the error. The RGI and $\overline{\text{MS}}$ results above refer to the three-flavour scheme. We compare the mesonic SU(3) LECs $X_0\in\{F_0, \Sigma_0, B_0\}$ with their SU(2) ChPT counterparts $X$, where the strange quark mass is fixed at its physical value, in the $\overline{\text{MS}}$ scheme with three active flavours at $2\;\GeV$: the decay constant $F_0<F\approx 86\;\MeV$~\cite{Aoki:2021kgd,ParticleDataGroup:2020ssz} and the chiral condensate $\Sigma_0<\Sigma\approx (270\;\MeV)^3$~\cite{Aoki:2021kgd} decrease significantly as we send the strange quark mass to zero, whereas the GMOR parameter $B_0\approx B\approx 2.66\;\GeV$ remains unaffected within its present uncertainty.

Further constraining the mass-dependence by including ensembles with lighter pion masses would be very interesting, in particular regarding the axial couplings. In addition to this, in the near future we plan to extend the analysis to the $N_f = 2+1$ case in order to further improve the accuracy, to test the applicability range of SU(3) ChPT and also to determine higher order LECs.

\begin{acknowledgments}
  The work of G.B.\ and S.W.\ is funded by the German Federal Ministry of Education and Research (BMBF) grant no.~05P18WRFP1. Additional support from the European Union’s Horizon 2020 research and innovation programme under the Marie Sk{\l}odowska\nobreakdash-Curie grant agreement no.~813942 (ITN EuroPLEx) and grant agreement no.~824093 (STRONG~2020) is gratefully acknowledged, as well as initial stage funding through the German Research Foundation (DFG) collaborative research centre SFB/TRR\nobreakdash-55.

  The authors gratefully acknowledge the \href{https://www.gauss-centre.eu}{Gauss Centre for Supercomputing (GCS)} for providing computing time through the \href{http://www.john-von-neumann-institut.de}{John von Neumann Institute for Computing (NIC)} on the supercomputer JUWELS~\cite{juwels} and in particular on the Booster partition of the supercomputer JURECA~\cite{jureca} at \href{http://www.fz-juelich.de/ias/jsc/}{J\"ulich Supercomputing Centre (JSC)}. GCS is the alliance of the three national supercomputing centres HLRS (Universität Stuttgart), JSC (Forschungszentrum Jülich), and LRZ (Bayerische Akademie der Wissenschaften), funded by the BMBF and the German State Ministries for Research of Baden\nobreakdash-Württemberg (MWK), Bayern (StMWFK) and Nordrhein\nobreakdash-Westfalen (MIWF). Additional simulations were carried out on the QPACE~3 Xeon Phi cluster of SFB/TRR\nobreakdash-55 and the Regensburg Athene~2 Cluster. The authors also thank the JSC for their support and for providing services and computing time on the HDF Cloud cluster~\cite{hdfcloud} at JSC, funded via the Helmholtz Data Federation (HDF) programme.

  Most of the ensembles were generated using \href{https://luscher.web.cern.ch/luscher/openQCD/}{\sc openQCD}~\cite{Luscher:2012av} within the \href{https://wiki-zeuthen.desy.de/CLS/}{Coordinated Lattice Simulations (CLS)} effort. We thank all our CLS colleagues for the joint generation of the gauge field ensembles. A few additional ensembles were generated employing the {\sc BQCD}\nobreakdash-code~\cite{Nakamura:2010qh} on the QPACE supercomputer of SFB/TRR\nobreakdash-55. For the computation of hadronic two- and three-point functions we used a modified version of the {\sc Chroma}~\cite{Edwards:2004sx} software package along with the {\sc Lib\-Hadron\-Analysis} library and the multigrid solver implementation of Refs.~\cite{Heybrock:2015kpy,Georg:2017zua} (see also ref.~\cite{Frommer:2013fsa}). We used {\sc Matplotlib}~\cite{Hunter:2007} to create the figures.
\end{acknowledgments}

\appendix
\section{FURTHER CHPT EXPRESSIONS\label{sec:ga_fv}}
\label{sec:axialadd}
We collect ChPT expressions that were not used in the final analysis. In particular, these are expressions that include decuplet loops (and therefore additional LECs that we were unable to resolve) and the finite volume effects for the axial charges. Regarding the latter, these have been computed using SU(2) HBChPT~\cite{Beane:2004rf} and confirmed in SU(2) BChPT~\cite{Khan:2006de}. We define the function
\begin{equation}
  h_1(\lambda_\pi)=
\sum_{\mathbf{n}\neq\mathbf{0}}\left[K_0(\lambda_\pi
  |\mathbf{n}|)-\frac{K_1(\lambda_\pi
  |\mathbf{n}|)}{\lambda_\pi|\mathbf{n}|}\right],
\label{eq:h2}
\end{equation}
that corresponds to $\mathbf{F_1}$ of Ref.~\cite{Beane:2004rf} while for $h(\lambda_\pi)$, defined in Eq.~\eqref{eq:h1}: $h(\lambda_\pi)=-(8/3)\mathbf{F_3}(M_\pi,L)$. Again $\lambda_{\pi}=LM_{\pi}$. The SU(3) finite size effects in the flavour symmetric limit (utilizing the couplings that are tabulated in Ref.~\cite{Ledwig:2014rfa} and truncating at $\mathcal{O}(p^3)$) read:
\begin{align}
  g_A^N(L)&=g_A^N-\frac{3}{2}(D+F)\xi h(\lambda_\pi)\nonumber\\&\,+\frac{2}{9}\left(27D^3+25D^2F+45DF^2+63F^3\right)\xi h_1(\lambda_\pi),
  \label{eq:fse1}\\
  g_A^{\Sigma}(L)&=g_A^{\Sigma}-3F\xi h(\lambda_\pi)+\frac{4}{9}F\left(25D^2+63F^2\right)\xi h_1(\lambda_\pi).\label{eq:fse2}
\end{align}

The gap between the decuplet and octet baryon mass in the chiral limit $\Delta=m_{D0}-m_0$ is within the range covered by our pion masses. Therefore, decuplet loop effects may in principle be relevant. Indeed, neglecting such terms, the finite volume effects of $g_A^B$ have a sign opposite to what we see in the data. Already in  Ref.~\cite{Jenkins:1991es} corrections due to transitions to decuplet baryons were considered. The full SU(3) result~\cite{MartinCamalich:2010fp} for the octet baryon mass for the case $m_s=m_{\ell}$, to be added to Eq.~\eqref{eq:mb}, reads:\footnote{For the LEC $\mathcal{C}$ we use the
  normalization of Refs.~\cite{Jenkins:1991es,WalkerLoud:2004hf,Beane:2011pc},
  where $\mathcal{C}^2=g^2_{\Delta N\pi}$~\cite{Beane:2004tw}.}
\begin{align}
m_B  &\mapsto m_B-\frac{\Delta^3}{(4\pi F_0)^2}\frac{5}{3}\mathcal{C}^2\left[\left(2-3\frac{M_\pi^2}{\Delta^2}\right)\log
    \left(\frac{M_\pi}{2\Delta}\right)\right.\nonumber\\&\qquad+\left.\frac{M_\pi^2}{2\Delta^2}+2\left(1-\frac{M_\pi^2}{\Delta^2}\right)w\left(\frac{M_\pi}{\Delta}\right)\right],\label{eq:barc}\\
    w(r)&=\left\{
  \begin{array}{ccc}
  {-\left(r^2-1\right)}^{1/2}\arccos\left(r^{-1}\right)&,&r\geq 1\\
  {\left(1-r^2\right)}^{1/2}\log\left(r^{-1}+\sqrt{r^{-2}-1}\right)&,&r< 1
  \end{array}\right.\label{eq:ccc}
\end{align}
with the additional LECs $\mathcal{C}$ and $\Delta$. Regarding the above decuplet baryon effects, we restrict ourselves to the heavy baryon approximation. The full EOMS BChPT result can be found in Ref.~\cite{MartinCamalich:2010fp}. Note that the decuplet decouples as $M_\pi\rightarrow 0$ as it should since in this case the extra term is proportional to $[3-4\log(M_\pi/(2\Delta))]M_\pi^4/(\Delta\, F_0^2)$, which is of a higher order in the chiral expansion. The associated finite volume corrections to Eq.~\eqref{eq:mb_fv} read~\cite{Procura:2006bj,Geng:2011wq}
\begin{widetext}
\begin{align}
  m_B(L)\mapsto
  m_B(L)+\frac{5}{3}\mathcal{C}^2\xi\frac{m_0^3}{(m_0+\Delta)^2}
\int_0^{\infty}\!\!\!\dd{y}\,\left\{\left(2-y+\frac{\Delta}{m_0}\right)f(y)
  \sum_{\mathbf{n}\neq\mathbf{0}}\left[f(y)K_0(\lambda_\pi|\mathbf{n}|f(y))
    -\frac{K_1(\lambda_\pi|\mathbf{n}|f(y))}{\lambda_\pi|\mathbf{n}|}\right]
\right\},\label{eq:barc2}
\end{align}
\end{widetext}
where
\begin{align}
  f(y)&=\sqrt{1+M_\pi^{-2}\left[\left(\Delta^2+2m_0\Delta-M_\pi^2\right)y
    +m_0^2y^2\right]}.
\end{align}
We refer to Ref.~\cite{MartinCamalich:2010fp} for the full SU(3) result and to Refs.~\cite{Procura:2006bj,Geng:2011wq} for the corresponding finite volume corrections.

  For the axial charges, we start from Ref.~\cite{Beane:2004rf} and implement the decoupling constraints~\cite{Bernard:1998gv,Hemmert:2003cb} at $\mathcal{O}(p^3)$. We obtain for the special $N_f=3$ case $m_s=m_{\ell}$:
\begin{align}
  g_A^B&\mapsto g_A^B-j_B\frac{\Delta^2}{16\pi^2F_0^2}J(M_\pi/\Delta)\nonumber\\
  &\qquad -n_B\frac{\Delta^2}{16\pi^2F_0^2}N(M_\pi/\Delta),
\end{align}
where
\begin{align}
  J(r)&=-r^2-\left(2-r^2\right)\log\left(\frac{r}{2}\right)-2w(r),
  \\
  N(r)&=-\frac{r^2}{3}+\frac{\pi r^3}{3}
  -\left(\frac23-r^2\right)\log\left(\frac{r}{2}\right)\nonumber\\
  &\qquad-\frac{2}{3}\left(1-r^2\right)w(r)
\end{align}
and the coefficients are given as
\begin{align}
  j_N&=5\left(F+D+\frac{8}{27}\mathcal{H}\right)\mathcal{C}^2,\\
  n_N&=-4\left(\frac{11}{9}D+F\right)\mathcal{C}^2,\\
  j_{\Sigma}&=10\left(F+\frac{5}{27}\mathcal{H}\right)\mathcal{C}^2,\\
  n_{\Sigma}&=-\frac{40}{9}D\,\mathcal{C}^2.
\end{align}
We remind the reader that the term $d_B\xi^{3/2}$ within Eqs.~\eqref{eq:d1} and~\eqref{eq:d2} does not appear at $\mathcal{O}(p^3)$ in the chiral expansion but is purely phenomenological. However, the function $\Delta^2 N(M_{\pi}/\Delta)$ contains a genuine term $\propto\xi^{3/2}(4\pi F_0)/\Delta$, justifying the inclusion of that parameter.

Regarding finite volume effects, we infer from Ref.~\cite{Beane:2004rf} (see also Ref.~\cite{Khan:2006de}) that the following terms need to be added to Eqs.~\eqref{eq:fse1}--\eqref{eq:fse2}
\begin{align}
  g_A^B(L)&\mapsto g_A^B(L)+
  \frac{4}{3}\xi\left[j_N\mathbf{F_2}(L M_\pi,M_\pi/\Delta)\right.\nonumber\\
&\qquad-\left.\frac{9}{8}n_N\mathbf{F_4}(L M_\pi,M_\pi/\Delta)\right],
  \label{eq:fsed}
\end{align}
where $\mathbf{F_2}$ and $\mathbf{F_4}$ are defined in Ref.~\cite{Beane:2004rf}.

\section{MODEL AVERAGING\label{sec:error}}

To address systematic effects we carry out fits varying the fit function (e.g., BChPT vs.\ HBChPT) as well as the number of data points included. This gives us a set of $N_M$ different results, one for each model $j$, from which we compute an average and its uncertainty that includes the statistical error and the systematic uncertainty due to the model variation. 

One widely used approach is to assign a weight $w_j$ given by the Akaike information criterion (AIC)~\cite{Akaike:1998zah} to each model~$j$ in the model averaging procedure. Here we employ the weights 
\begin{align}
  w_j= A \exp\left\{ -\frac{1}{2} \Big[ \text{max}\;(\chi^2_{j},N_{\text{dof},j}) - N_{\text{dof},j} + k_{j} \Big] \right\},
\end{align}
see, e.g., Eq.~(161) of the e-print version of Ref.~\cite{Borsanyi:2020mff} and references therein.\footnote{Recently, instead of subtracting $N_{\text{dof}}-k$ from $\chi^2$ in the exponent, in Ref.~\cite{Jay:2020jkz} it has been suggested to subtract $-2n_{\text{cut}}-2k=\text{const}+2N_{\text{dof}}$, where $n_{\text{cut}}$ is the number of removed data points. This seems counter-intuitive: since for a good fit $\chi^2\sim N_{\text{dof}}$, this change would result in a very strong preference for fits that include as many data points as possible, even if the corresponding $\chi^2/N_{\text{dof}}$-values were significantly larger.} The normalization $A$ is such that $\sum_i^{N_M} w_i = 1$. $\chi^2_{j}$ denotes the $\chi^2$-value of the fit to model $j$, $k_{j}$ the number of fit parameters and $N_{\text{dof},j} = n_j - k_j$ the number of degrees of freedom. By replacing $\chi^2\mapsto\max(\chi^2,N_{\text{dof}})$, we deviate somewhat from Ref.~\cite{Borsanyi:2020mff} in so far as reducing the $\chi^2$-value below $N_{\text{dof}}$ will not further increase the weight. The rationale for this choice is that if the fit function perfectly described the data then a value $\chi^2<N_{\text{dof}}$ should not be more likely than the expected value $\chi^2=N_{\text{dof}}$. The above equation extends the AIC to also varying the number of data points $n_j$ and not only the fit function. It is valid as long as there are no correlations between the removed and the remaining data points, the fit function is smooth and the parametrization does not depend on the data space. This applies to our case where we reduce the number of data points by removing entire ensembles and carry out the same set of fits for every data set.

For each parameter $a$ that we are interested in, we generate for each model $j$ a bootstrap distribution $a_j(b)$ with $N_b = 500$ bootstrap samples $b$. The (normalized) bootstrap histograms are usually normal distributed,
\begin{align}
  f_j(a) = \frac{1}{\sqrt{2\pi}\sigma_{j}} \exp{\left\{ -\frac{1}{2} \left( \frac{a-a_j}{\sigma_j} \right)^2 \right\}},
\end{align}
with a mean $a_j$ and a standard deviation $\sigma_j$. From the (discrete) histograms, we obtain the model averaged distribution
\begin{align}
  f(a) = \sum_j w_j f_j(a),
\end{align}
from which we take the median and the $1\sigma$ confidence interval determined by the 15.9\% and 84.1\% percentiles as the model average $\bar{a}$ and its upper and lower confidence limits $\bar{a}+\Delta a_+$ and $\bar{a}-\Delta a_-$. We then quote the average and its total error as $\bar{a}_{\Delta a_-}^{\Delta a_+}$. This procedure is illustrated in Figs.~\ref{fig:f0b0_weighted} and~\ref{fig:m0ga_weighted}, where the histograms are coarsely binned for a better visualization.
\end{document}